\documentstyle[epsfig]{mn}

\def\gs{\mathrel{\raise0.35ex\hbox{$\scriptstyle >$}\kern-0.6em
\lower0.40ex\hbox{{$\scriptstyle \sim$}}}}
\def\ls{\mathrel{\raise0.35ex\hbox{$\scriptstyle <$}\kern-0.6em
\lower0.40ex\hbox{{$\scriptstyle \sim$}}}}

\begin{document}

\title
[Hierarchical star formation/AGN fuelling] 
{Dust-obscured star formation and AGN fuelling in hierarchical models of 
galaxy evolution}
\author
[A.\,W. Blain et al.]
{
A.\,W. Blain,$^{1,2}$ Allon Jameson,$^1$ Ian Smail,$^{\! 3}$ M.\,S. Longair,$^{1}$ 
J.-P. Kneib$^2$ \and and R.\,J. Ivison$^{4}$\\
\vspace*{1mm}\\
$^1$ Cavendish Laboratory, Madingley Road, Cambridge, CB3 0HE, UK.\\
$^2$ Observatoire Midi-Pyr\'en\'ees, 14 Avenue E. Belin, 31400 Toulouse, France.\\
$^3$ Department of Physics, University of Durham, South Road, Durham,
DH1 3LE, UK.\\
$^4$ Department of Physics \& Astronomy, University College London, Gower 
Street, London, WC1E 6BT, UK.\\
}
\maketitle

\begin{abstract}
A large fraction of the luminous distant submillimetre-wave galaxies 
recently detected using the Submillimetre Common-User Bolometer 
Array (SCUBA) camera on the James Clerk Maxwell Telescope appear to be 
associated with interacting optical counterparts. We investigate the nature
of these systems using a simple hierarchical clustering model of galaxy 
evolution, in which the large luminosity of the SCUBA galaxies is 
assumed to be generated at the 
epoch of galaxy mergers in a burst of either star formation activity or the 
fuelling of an active galactic nucleus (AGN). The models are well 
constrained by the observed spectrum of the far-infrared/submillimetre-wave 
background radiation and the 60-$\mu$m counts of low-redshift 
{\it IRAS} galaxies. The ratio between the total 
amount of energy released during mergers and the mass of dark matter involved 
must increase sharply with redshift $z$ at $z \ls 1$, and then decrease at greater 
redshifts. This result is independent of the fraction of the luminosity of
mergers that is produced by starbursts and AGN. One additional parameter -- 
the reciprocal of the 
product of the duration of the enhanced luminosity produced by the merger 
and the fraction of mergers that induce an enhanced luminosity, which we call 
the activity parameter -- is introduced, to allow the relationship between 
merging dark matter haloes and the observed counts of distant dusty galaxies 
to be investigated. 
The observed counts can only be reproduced if the activity parameter is greater 
by a factor of about 5 and 100 at redshifts of 1 and 3 respectively, as 
compared with the present epoch. Hence, if merging galaxies 
account for the population of SCUBA galaxies, then the merger process must 
have been much more violent at high redshifts. We discuss the counts of 
galaxies and the intensity of background radiation in the optical/near-infrared 
wavebands in the context of these hierarchical models, and thus investigate 
the relationship between the populations of submillimetre-selected 
and Lyman-break galaxies.
\end{abstract}  


\section{Introduction}

The history of star formation in dusty galaxies was recently discussed by 
Blain et al. (1999c), who assumed that the 
distant galaxies recently detected 
using the 450/850-$\mu$m Submillimetre Common-User Bolometer Array 
(SCUBA) camera (Holland et al.\ 1999) were the high-redshift 
counterparts of local ultraluminous {\it IRAS} galaxies. 
The global star formation rate (SFR) in dust obscured galaxies was inferred to 
be significantly greater than 
that of optically selected high-redshift galaxies (Steidel et al.\ 
1996a,b, 1999), subject to the uncertain fraction of the luminosity of the 
submillimetre-selected samples of galaxies (Smail, Ivison \& Blain 1997; 
Barger et al.\ 1998; Hughes et al.\ 1998; Barger et al.\ 1999a; 
Blain et al.\ 1999b; Eales et al.\ 
1999) that is produced by accretion processes in active galactic nuclei (AGN). A 
fraction of at most 30\,per cent, and more likely 10--20\,per cent, is 
suggested by both 
follow-up observations (Frayer et al.\ 1998; Ivison et al.\ 1998; Smail et al.\ 1998; 
Barger et al.\ 1999b; Frayer et al.\ 1999; Lilly et al.\ 1999), and information derived 
in other wavebands; see section\,5.4 of Blain et al.\ (1999c), Almaini, Lawrence \& 
Boyle (1999) and Gunn \& Shanks (1999). Using a different approach, in which 
the high-redshift SCUBA population is decoupled from the local 
infrared-luminous galaxies, Trentham, Blain \& Goldader (1999) were able to 
reconcile the SCUBA counts with a less dramatic amount of obscured 
star-formation activity. Use another empirical approach, Tan, Silk \& Balland 
(1999) derived results somewhere between the two. 
A summary of the existing data on the history of star formation is presented 
in Fig.\,1. 

\begin{figure*}
\begin{minipage}{170mm}
\begin{center}
\epsfig{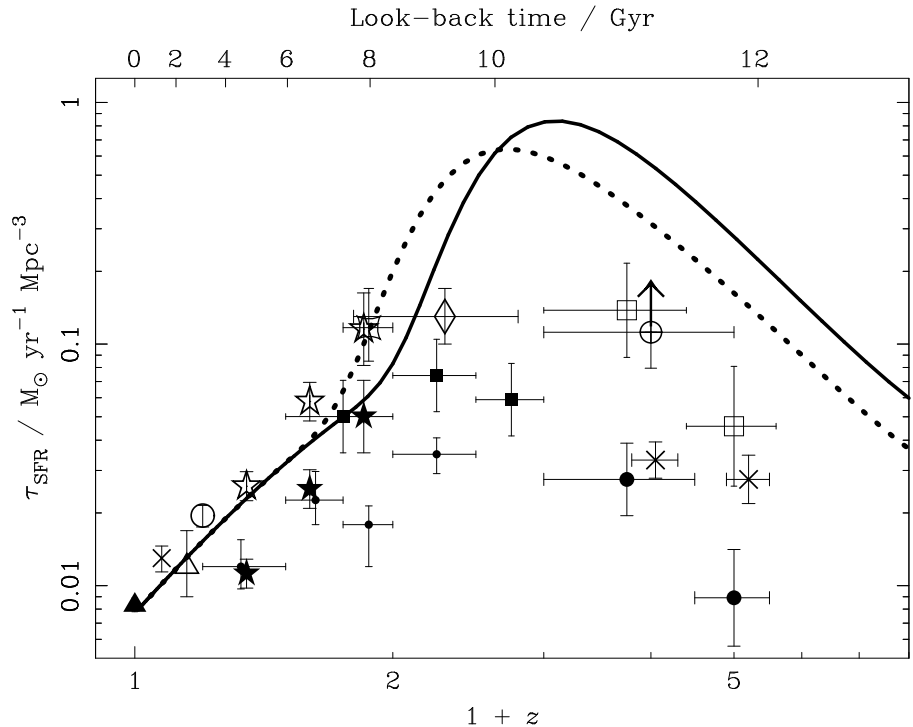}
\end{center}
\caption{The SFR as a function of redshift, $\tau_{\rm SFR}(z)$, as inferred from 
ultraviolet/optical/near-infrared observations, in order of increasing redshift, by 
Gallego et al.\ (1996; filled triangle), Gronwall (1999; thin diagonal cross), 
Treyer et al.\ (1998; open triangle), Tresse \& Maddox (1998; empty circle), Lilly 
et al.\ (1996; filled stars), Cowie, Songaila \& Barger (1999; small filled
circles); Glazebrook et al.\ (1999; bent square at $z=0.9$), Connolly et al.\ 
(1997; filled squares), Yan et al.\ (1999; empty lozenge), Madau et al.\ (1996; 
large filled circles), and Pettini et al.\ (1998a,b; empty squares). Flores et al.\ 
(1999; empty stars) and Pettini et al.\ (1998a,b) have corrected the 
Lilly et al. and Madau et al. results respectively for observed dust extinction. 
The high-redshift points are derived from analyses of the {\it Hubble Deep 
Field (HDF)}. A recent ground-based survey of a wider area by Steidel et al.\ 
(1999) increased the estimated high-redshift SFR, as shown by the thick 
diagonal crosses. No extinction correction is applied to these points; Steidel 
et al.\ (1999) estimate that the extinction-corrected SFR is greater by a 
factor of about 5. Other analyses of the {\it HDF} data by Sawicki, Lin \& Yee 
(1997) and Pascarelle, Lanzetta \& Fern\'andez-Soto (1998) yielded similar 
results. Cowie, Songaila \& Barger (1999) have argued recently that the 
low-redshift SFR increases more gradually, as $(1+z)^\gamma$ with 
$\gamma \simeq 1.5$, in contrast with the usual 
$\gamma \simeq 4$ (Lilly et al.\ 1996). At longer 
wavelengths, SFRs have been derived from mid-infrared observations by 
Flores et al.\ (1999) and from submillimetre-wave observations by Blain et al.\ 
(1999c; lines) and Hughes et al.\ (1998; empty circle with upward pointing 
arrow). Cram (1998) investigated 
the star-formation history at $z \ls 1$ using radio observations, deriving a 
rate of $2.5 \times 10^{-2}$\,M$_\odot$\,yr$^{-1}$\,Mpc$^{-3}$ at $z=0$, which 
increases by an order of magnitude at $z\simeq1$. 
The submillimetre-based Gaussian and 
modified Gaussian SFR models (see Blain et al.\ 1999c and Barger et al.\ 1999b) 
are represented by the solid and dotted lines respectively. A better fit to the 
preliminary redshift distribution of submillimetre-selected galaxies (Barger et al.\ 
1999b) is provided by the modified model, in which the parameters 
$H$, $\sigma$ and $z_{\rm p}$ (Blain et al.\ 1999c) take the values 
70, 1\,Gyr and 1.7 respectively, as compared with 95, 0.8\,Gyr and 2.1 in the 
original model. The total amount of star formation/AGN 
fuelling activity taking place in the modified Gaussian model is 90\,per cent of 
that in the original Gaussian model. 
}
\end{minipage}
\end{figure*}

Although well constrained, and in accord with the available observational data, 
the models in Blain et al.\ (1999c) and Trentham et al.\ (1999) included few 
details of the physical origin of the large luminosity of SCUBA galaxies. 
Semi-analytic models of hierarchical galaxy formation, in which galaxies 
assemble by the merger of progressively larger subunits (Cole et al. 1994; Baugh
et al.\ 1998; Kauffmann \& Charlot 1998; Somerville, Primack \& Faber 1999) 
have been used to account for a wide range of observations in the optical and 
near-infrared wavebands, and have been extended into the far-infrared and 
submillimetre wavebands by Guiderdoni et al.\ (1998). These models involve 
a large number of free parameters, and the
interplay between them can make it difficult to identify the most important
physics responsible for a particular observation. In this paper we 
develop a model of infrared-luminous galaxies in a simple 
version of such a scenario (Blain \& Longair 1993a,b; Jameson, Longair \& Blain 
1999), which includes many fewer parameters and hopefully makes the
astrophysics more transparent. We attempt to 
reproduce the SCUBA counts by invoking bright dust-enshrouded bursts of 
either star formation activity or AGN fuelling at the epochs of mergers. 

Motivation for considering the SCUBA galaxies as luminous mergers is provided 
by both the optical identifications of the Smail et al.\ (1998) sample, which 
appear to contain a large fraction of interacting galaxies, and the gas 
consumption rate that is inferred from observations of CO emission of two 
submillimetre-selected galaxies made using the Owens Valley Millimeter Array 
(Frayer et al.\ 1998, 1999), which cannot be sustained for more than a few 
$10^8$\,yr. 
Even the faint and compact counterparts listed in Smail et al.\ (1998) could 
be merging galaxies, but too faint to identify as such; see the simulations 
of the appearance of high-redshift mergers in Bekki, Shioya \& Tanaka (1999). 
If the SCUBA galaxies are the high-redshift counterparts of the 
low-redshift ultraluminous infrared galaxies, which are predominantly 
merging systems, then this also offers support for modeling the SCUBA 
galaxies as mergers. Using our simple model, we emphasise the most 
important features and the underlying physics of the evolution of 
submillimetre-selected galaxies and their relationship to the population of 
quiescent galaxies. 

In Section\,2 we describe the details of the model, and investigate the constraints 
imposed by the intensity of the far-infrared and submillimetre-wave background 
radiation and the counts of low-redshift {\it IRAS} galaxies. We discuss the 
evolution of the luminosity density in the model and compare the models 
with observations in the same way as the evolving {\it IRAS} luminosity 
function models discussed by Blain et al.\ (1999c). 
In Section\,3 we discuss the predictions in the 
context of source counts in the submillimetre and far-infrared wavebands, and 
investigate whether the SCUBA galaxies can easily be explained in an 
hierarchical picture of galaxy formation and evolution. In Section 4 the 
corresponding background radiation intensities and galaxy counts in the 
near-infrared and optical wavebands are discussed. In Section 5 we review the 
parameters we have introduced to describe the models. We present our 
main conclusions in Section 6. A value of Hubble's 
constant $H_0=100h$\,km\,s$^{-1}$\,Mpc$^{-1}$, with $h=0.5$, a density 
parameter $\Omega_0=1$ and a cosmological constant $\Omega_\Lambda=0$ 
are assumed. 

\section{An analytic hierarchical picture}

The evolution of galaxy-scale structures under gravity according to hierarchical 
clustering models can be analysed using the
Press--Schechter formalism (Press \& Schechter 1974), which describes the 
time-dependent mass spectrum of bound objects. The analytic  
results of the Press--Schechter formalism are in quite acceptable agreement with 
those of N-body simulations (Brainerd \& Villumsen 1992). 
The formalism can be extended to yield a very straightforward semi-analytic 
merger rate, under the single assumption that the process of halo mergers is 
independent of mass (Blain \& Longair 1993a,b). 

\subsection{The Press--Schechter Formalism} 

According to the Press--Schechter prescription, the mass spectrum of 
bound objects with masses between $M$ and 
$M+{\rm d}M$ is
\begin{equation}
N_{\rm PS}(M,z) = { {\bar\rho} \over {\sqrt \pi} } {\gamma \over {M^2} } 
\left( { M \over {M^*} } \right)^{\gamma/2} 
\exp\left[ - \left( { M \over {M^*} } \right)^\gamma \right], 
\end{equation}  
in which $\bar\rho$ is the smoothed comoving density of the Universe, dominated by dark 
matter, $\gamma = (3+n)/3$, where $n$ is the power-law index of primordial 
density fluctuations, and $M^*(z)$ is a parameter which describes the evolution 
of density fluctuations as a function of redshift $z$:  
\begin{equation}
M^*(z) = M^*(0) \left[ { {\delta(z)} \over {\delta(0)} } \right]^{2/\gamma}. 
\end{equation}
$\delta(z)$ is the function describing the growth of perturbations in a general 
cosmology, and is derived from the equation,  
\begin{equation}
{ \ddot\delta } + 2 { {\dot R} \over R } \dot\delta - 
{ {4\pi G \bar \rho} \over {R^3} }  \delta = 0,
\end{equation} 
in which $R$ is the scalefactor of the Universe. In an Einstein de-Sitter model, 
the growing mode has $\delta \propto (1+z)^{-1}$. $M^*(0)$ is the typical mass 
of bound objects at $z=0$.
Inhomogeneities do not grow if $\gamma \le 0$, that is if $n \le -3$. For 
scale-invariant density fluctuations, $n=1$ or $\gamma = 4/3$. This is close to 
the value observed on the largest scales from the cosmic microwave background 
radiation (CMBR). Observations of large-scale structure indicate that 
$n \simeq -1.5$, or $\gamma \simeq 1/2$, on the smallest scales 
(Peacock \& Dodds 1994), which can be associated with the transfer function 
between a primordial $n=1$ spectrum and the spectrum after recombination. 

Using the Tully--Fisher relation (Hudson et al. 1998), 
Blain, M\"oller \& Maller (1999) obtained a value 
$M^*(0)=3.6\times10^{12}$\,M$_\odot$. The exact value of $M^*(0)$ is not 
very important here, as a mass-to-light ratio is introduced to 
convert the mass spectrum into a luminosity function. 

\subsection{Deriving a merger rate} 

Working from the mass spectrum $N_{\rm PS}(M,z)$, 
Blain \& Longair (1993a,b) showed that a formation  
rate of bound objects $\dot N_{\rm form}(M,z)$ in galaxy halo mergers can be 
constructed if the mass distribution of the components involved in a statistical 
sample of merger events is assumed to be independent of mass. In this case, the 
merger rate can be represented accurately by the function
\begin{equation} 
\dot N_{\rm form} = \dot N_{\rm PS} + \phi { {\dot M^*} \over {M^*} } N_{\rm PS} 
\exp\left[ (1-\alpha) \left( { M \over {M^*} } \right)^\gamma \right],   
\end{equation}  
where 
\begin{equation}
\dot N_{\rm PS} = \gamma { {\dot M^*} \over {M^*} } N_{\rm PS} 
\left[ \left( { M \over {M^*} } \right)^\gamma - { 1 \over 2 } \right]. 
\end{equation}
$\phi$ and $\alpha$ are numerical constants, typically about 1.7 and 1.4; 
their exact values depend on the assumed mass distribution of merging 
components (Blain \& Longair 1993a), but have little effect on the results. 
The values of both $\phi$ and $\alpha$ are weak functions of $\gamma$ and 
depend on the world model parameters, 
but the form $\phi/\sqrt{\alpha}$ that appears in the
calculations of the background radiation intensity and metal abundance 
is almost independent of the value of $\gamma$.

\subsection{Deriving observable quantities} 

The merger rate as a function of mass $\dot N_{\rm form}$ can be readily used to 
estimate a number of observable quantities, starting with the luminosity 
density (or volume emissivity), 
\begin{equation} 
\epsilon_{\rm L}(z) = 0.007 c^2 { {x(z)} \over { 1 - f_{\rm A} } } 
\int M \dot N_{\rm form} \, {\rm d}M, 
\end{equation}
in which $x(z)$ is the ratio of the mass of baryonic matter converted into 
metals by nucleosynthesis in a merger-induced starburst to the total dark mass 
involved in the merger. The 
rationale behind this form of relation is given by Longair (1998). The factor of 
0.007 is the approximate efficiency of conversion of mass into energy in stellar
nucleosynthesis. The parameter $f_{\rm A} < 1$ describes the fraction of the 
total luminosity of merging galaxies that is attributable to accretion in AGN, and 
is expected to lie in the range $0.1 \le f_{\rm A} \le 0.3$ (Genzel et al.\ 1998; 
Lutz et al.\ 1998; Almaini et al.\ 1999; 
Barger et al.\ 1999b; Gunn \& Shanks 1999).
The parameter $x(z)$ is expected to vary with redshift $z$. Blain \& Longair 
(1993b) predicted a flat background spectrum in the submillimetre and 
far-infrared wavebands, assuming a constant value of 
$x$. Subsequent observations (e.g. Fixsen et al.\ 1998) demand a 
redshift-dependent form of $x(z)$, as discussed by Blain et al.\ (1999c). 

By evaluating the integral in equation (6), the luminosity density can be 
expressed as
\begin{equation} 
\epsilon_{\rm L}(z) = 0.007 c^2 { {x(z)} \over { 1 - f_{\rm A} } } 
\bar\rho { \phi \over {\sqrt{\alpha}} } { {\dot M^*} \over {M^*} }. 
\end{equation}
Interestingly,
\begin{equation}
{ {\dot M^*} \over {M^*} } = { 2 \over \gamma } { {\dot \delta(z)} \over {\delta(z)}}, 
\end{equation}
and so, because the density contrast $\delta(z)$ is not a function of
the perturbation spectral index $\gamma$, the $\gamma$ dependence in this 
term is just a simple scaling. Thus the effect of the value of $\gamma$ on the 
background radiation intensity can be studied or removed very easily. In an 
Einstein--de Sitter model the luminosity density 
$\epsilon_{\rm L} \propto x(z) (1+z)^{3/2}$. 

The comoving density of metals produced in starbursts between a redshift
$z_0$, at which star formation activity begins, and $z$, is
\begin{equation}
\rho_{\rm m}(z) = \bar\rho { \phi \over {\sqrt{\alpha}} } \int_z^{z_0} 
{ 1 \over c} { {x(z)} \over {(1+z)}} { {\dot M^*} \over {M^*} } 
{ {{\rm d}r} \over {{\rm d}z}} \, {\rm d}z, 
\end{equation} 
where $r$ is the radial comoving distance coordinate. Note that this result
depends on the merger efficiency parameter $x$ but not on the AGN fraction 
$f_{\rm A}$, as metals are only generated in merger-induced 
starbursts and not in AGN fuelling events. Similarly, the 
background radiation 
intensity per unit solid angle emitted by these galaxies, which have a 
spectral energy distribution (SED) $f_\nu$, is 
\begin{equation} 
I_\nu = { 1 \over {4 \pi} } \int_0^{z_0} { { \epsilon_{\rm L}(z) } \over {1+z} } 
\,{ {f_{\nu(1+z)}} \over { \int f_{\nu'} \,{\rm d}\nu' } } 
{ {{\rm d}r} \over {{\rm d}z}} \, {\rm d}z. 
\end{equation} 
If the form of $\epsilon_{\rm L}$ (equation 7) is included explicitly, then 
\begin{equation}
I_\nu = { {0.007c^2 \bar\rho} \over {4 \pi (1-f_{\rm A})} } 
{ \phi \over {\sqrt{\alpha}} } \int_0^{z_0} { {x(z)} \over {1+z} } 
{ {\dot M^*} \over {M^*} } 
{ {f_{\nu(1+z)}} \over { \int f_{\nu'} \,{\rm d}\nu' } } 
{ {{\rm d}r} \over {{\rm d}z}} \, {\rm d}z. 
\end{equation} 
Details of the assumed dust SED can be found in Blain et al.\ (1999c). The
mid-infrared SED is assumed to take the form $f_\nu \propto \nu^{-1.7}$. 

None of the quantities listed above are affected by the time dependence of the 
release of energy during merging events; however, the source count 
requires the time profile of the merger induced 
starburst/AGN to be included. For simplicity, this profile is assumed to have a 
top-hat form with duration $\sigma$. The time profile of the luminosity generated
in a detailed simulation of the merger of gas-rich galaxies is discussed by 
Mihos \& Hernquist (1996), Bekki et al.\ (1999)
and Mihos (1999).  
The typical duration of AGN fuelling 
events and starbursts may differ; for example, a lower limit to the duration of 
a starburst is set by the lifetime of the highest mass stars, but there is no 
lower limit to the duration of an AGN fuelling event. However, to avoid 
introducing an unnecessarily complicated model, the time-scale of a merger 
induced luminous 
phase is assumed to be independent of its origin. In addition, because not all 
the mergers of dark matter haloes that take place at each epoch need 
induce a starburst/AGN, a fraction $F \le 1$ is assumed. Again, this fraction 
could differ for starbursts and AGN, but for 
simplicity it is assumed not to. The luminosity of a typical merger induced 
starburst/AGN of mass $M$ is thus
\begin{equation} 
L(M, z) = 0.007c^2 { {x(z)} \over {1-f_{\rm A}} } { 1 \over {F \sigma} } M. 
\end{equation} 
The source count $N$ of galaxies per unit solid angle brighter than a flux 
density $S_\nu$ is  
\begin{equation} 
N(S_\nu) = \int_0^{z_0} \int_{M_{\rm min}}^\infty F \sigma 
\dot N_{\rm form}(M, z) \,{\rm d}M D^2(z) \, { {{\rm d}r} \over {{\rm d}z}} \,
{\rm d}z, 
\end{equation} 
where $D(z)$ is the comoving distance parameter. 
The minimum mass merger visible at a flux density $S_\nu$ and redshift $z$ is 
\begin{equation} 
M_{\rm min} = { {4 \pi D^2 (1+z) S_\nu } \over { 0.007c^2 } } 
{ {F \sigma} \over { x(z) } } ( 1 - f_{\rm A}) 
{ {\int f_{\nu'}\,{\rm d}\nu'} \over {f_{\nu(1+z)}} }. 
\end{equation} 
The time-scale and bursting fraction parameters, $\sigma$ and $F$, always
appear together in calculations, and thus the single parameter $F\sigma$
is constrained by observations. We define $(F\sigma)^{-1}$ to be the 
activity parameter, which is large in violent starbursts/AGN and free to 
vary as a function of redshift. 
The rate of energy release within each 
individual starburst/AGN is controlled by the 
value of the activity parameter. Within a representative cosmological volume 
the presence of a population of either rare long-lived or common short-lived 
starbursts/AGN cannot be distinguished. This is why the time-scale and 
bursting fraction parameters $\sigma$ and $F$ are bound together in the 
activity parameter. 

Incorporating redshift evolution of the activity 
parameter introduces another degree of freedom into the count model, in 
addition to that provided by the redshift evolution of the star formation/AGN 
fuelling efficiency parameter $x$. 
Of course, $x$, $F$ and the time-scale $\sigma$ are also free 
to vary as a function of mass. At present, we find no compelling reason to 
incorporate this additional complication into the model.

\begin{figure*}
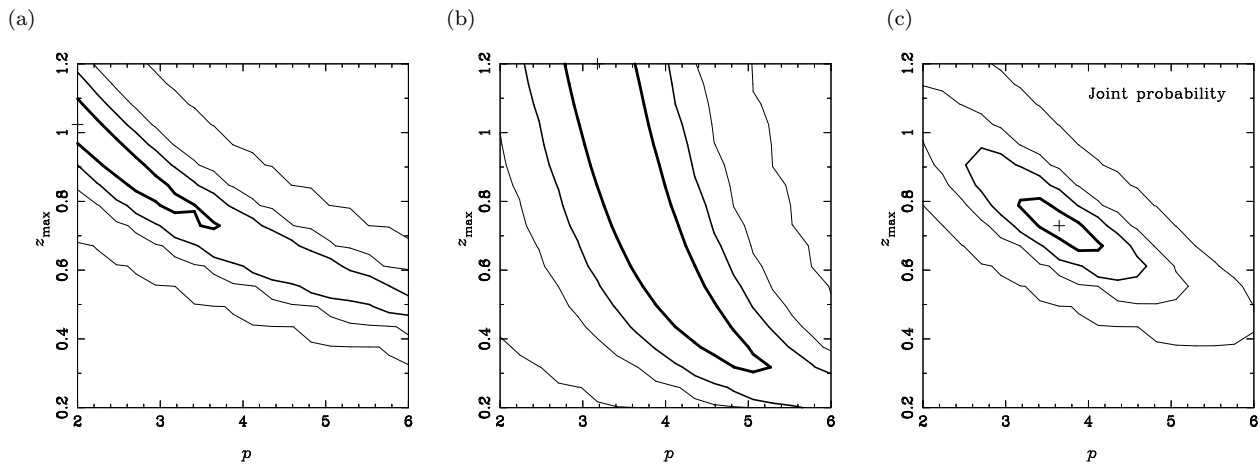

\begin{minipage}{170mm}
(a) \hskip 53.5mm (b) \hskip 53.5mm (c)
\begin{center}
\vskip -2mm
\epsfig{file=fig2a.ps, width=5.45cm, angle=-90} \hskip 5mm
\epsfig{file=fig2b.ps, width=5.45cm, angle=-90} \hskip 5mm
\epsfig{file=fig2c.ps, width=5.45cm, angle=-90}
\end{center}
\caption{The results of fitting (a) the background radiation spectrum, (b) the 
shape of the {\it IRAS} 60-$\mu$m counts (see Fig.\,4a for references to the 
data) and (c) both sets of data. The contours are drawn 
1, 2, 3 and 5$\sigma$ away from the peak probability, which is marked by a cross. 
This example is calculated for $T_{\rm d}=45$\,K. 
}
\end{minipage}
\end{figure*}

\subsection{Constraining the parameters} 

The background radiation intensities and source counts calculated from the 
equations derived above depend on a range of parameters: the world 
model, defined by $H_0$, $\Omega_0$ and $\Omega_\Lambda$; the perturbation 
spectral index $n$ and the value of $M^*(0)$; the constants $\phi$ and $\alpha$ 
in the merger rate; the merger efficiency $x(z)$; the AGN fraction $f_{\rm A}$; 
the fraction of mergers that lead to a starburst/AGN $F$; their duration $\sigma$; 
and their SED $f_\nu$.  

\begin{table}
\caption{Values of the parameters $p$ and $z_{\rm max}$ required in the 
expression for the merger efficiency $x(z)$ (equation 16) to fit the 
submillimetre/far-infrared background radiation spectrum and the
low-redshift 60-$\mu$m 
{\it IRAS} galaxy counts as a function of assumed dust temperature $T_{\rm d}$. 
The corresponding values of $x_0$,  $F\sigma$ and $\Omega_{\rm m}$ are 
also listed. $F\le 1$. The star formation histories associated with these models
are shown in Fig.\,7. Note that the most intense star formation activity 
takes place at a redshift $z \simeq 5 z_{\rm max}$. 
}
{\vskip 0.75mm}
{$$\vbox{
\halign {\hfil #\hfil && \quad \hfil #\hfil \cr
\noalign{\hrule \medskip}
$T_{\rm d}$/K & $p$ & $z_{\rm max}$ & $x_0 / \gamma (1-f_{\rm A})$ &
($F\sigma/\gamma$)/Gyr & 
$\Omega_{\rm m}/ 10^{-3}$ \cr
& & & & (at $z=0$) & $ \times (1-f_{\rm A})$ \cr
\noalign{\smallskip \hrule \smallskip}
35 & %5.3 
4.4 & 0.44 & $4.32 \times 10^{-5}$ & 0.040 & $1.2$ \cr %\times 10^{-3}$ \cr
40 & %4.6 
4.0 & 0.55 & $4.08 \times 10^{-5}$ & 0.041 & $1.3$ \cr %\times 10^{-3}$ \cr
45 & %3.6 
3.3 & 0.73 & $4.68 \times 10^{-5}$ & 0.063 & $1.4$ \cr %\times 10^{-3}$ \cr
50 & %2.7 
2.7 & 0.96 & $6.36 \times 10^{-5}$ & 0.141 & $1.7$ \cr %\times 10^{-3}$ \cr
\noalign{\smallskip \hrule}
\noalign{\smallskip}\cr}}$$}
\end{table}

Blain et al.\ (1999c) used the low-redshift 60-$\mu$m {\it IRAS} source count and 
the 175-$\mu$m {\it ISO} counts to constrain their models; here, however, we 
use the bright 60-$\mu$m counts and the form of the far-infrared/submillimetre 
background spectrum, the two best determined observables, to constrain the 
parameters that describe the merger efficiency $x(z)$. If the SED does not 
depend on the mass of the merging galaxies, then the background spectrum 
(equation 10) is determined entirely by the form of 
the merger efficiency $x(z)$, 
\begin{equation} 
I_\nu \propto {\phi \over {\gamma \sqrt{\alpha}} } \int_0^{z_0} 
{ {x(z)} \over {1-f_{\rm A}} }  
{ {\dot \delta(z)} \over {\delta(z)}} { 1 \over {1+z} } 
{ {f_{\nu(1+z)}} \over {\int f_{\nu'}\,{\rm d}\nu'} } 
{ {{\rm d}r} \over {{\rm d}z}} \, {\rm d}z. 
\end{equation}
Note that the dependence of $I_\nu$ on the perturbation index $n$ through 
$\gamma$ is completely separate from the dependence on the world model. 
Thus the background spectrum determined by Puget et al. (1996), 
Guiderdoni et al.\ (1997), Dwek et al.\ (1998), Fixsen et al.\ (1998), Hauser et al.\ 
(1998), Schlegel, Finkbeiner \& Davis (1998) and Lagache et al.\ (1999) can 
always be used to constrain the form of the merger efficiency $x(z)$. We adopt 
a form of $x(z)$ identical to the `peak model' 
described in Blain et al. (1999c):
\footnote{Note that the form of this equation published in 
Blain et al.\ (1999c)
contained a typographical error in the index of $(1+z)$.} 
\begin{equation} 
x(z) = \displaystyle{ 2 x_0  \left[ 1 + \exp{ { z \over {z_{\rm max}} } } \right ]^{-1}
(1+z)^{p+(2 z_{\rm max})^{-1}}. }
\end{equation}
This is not a uniquely 
appropriate functional form of $x(z)$. It was originally chosen to allow the 
star-formation history derived by Madau et al.\ (1996) to be fitted. Its 
three parameters can be manipulated to produce a wide range of 
plausible star formation histories. The three parameters are: $p$, 
the asymptotic low-redshift slope of the merger efficiency $x(z)$ in 
$(1+z)$; $z_{\rm max}$, the redshift above which the high-redshift exponential 
cut-off starts to take effect; and $x_0$ the value of $x(0)$. The epoch of 
most intense star-formation/AGN-fuelling corresponds to a 
redshift $z \simeq 5 z_{\rm max}$ in these models.

In Figs\,2(a) and (b) the probabilities of fitting the 
background radiation spectrum and the slope of the 60-$\mu$m counts 
predicted from the merger efficiency $x(z)$, defined in equation (16), 
to observations are shown as a function of the key parameters $p$ and 
$z_{\rm max}$, as an example for a dust temperature of 45\,K. In Fig.\,2(c) 
the joint probability of 
fitting both sets of data is shown. Note that a constant dust temperature is 
assumed. The value of $f_{\rm A}$ does not affect the results. 
If different dust 
temperatures are assumed, then the position of maximum probability moves 
around the $p$--$z_{\rm max}$ plane. However, when the form of the evolution 
of luminosity density is calculated for each temperature, the curve 
has a similar form. The best-fitting values of $p$ and $z_{\rm max}$, and the 
corresponding values of the merger efficiency $x_0$, the activity
parameter $(F\sigma)_0^{-1}$ and the density parameter in metals at $z=0$, 
$\Omega_{\rm m}$, are presented in Table\,1 for four plausible values of the 
dust temperature: 35, 40, 45 and 50\,K. Note that the best fitting values of 
$p$ and $z_{\rm max}$ depend only weakly on the world model assumed. 

The physical processes that demand a form of luminosity density which 
rises steeply with increasing redshift before turning over have not been 
considered here in any detail. It seems likely, however, that the steep
decline in the star formation rate to the present day is related to the 
declining gas content of galaxies at $z \ls 1$, and that the behaviour at high
redshifts could be attributable to relatively inefficient cooling of gas and 
thus of star formation in the mergers of metal-poor high-redshift systems (see 
Pei \& Fall 1995 and Pei, Fall \& Hauser 1999 for discussions of gas and dust 
evolution in the Universe). We discuss these issues further in Jameson, 
Blain \& Longair (2000). 

\begin{figure}
\begin{center}
\epsfig{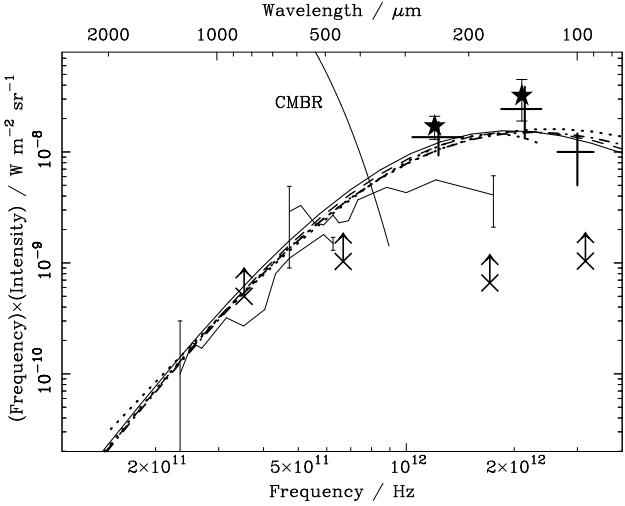}
\end{center}
\caption{The intensity of background radiation in the millimetre, submillimetre
and far-infrared wavebands, as deduced by: Puget et al.\ (1996) -- thin solid lines
with error bars at the ends; Fixsen et al.\ (1998) -- thin dotted line that ends
within the frame; Schlegel et al.\ (1998) -- stars; Hauser et al.\ (1998) and Dwek 
et al.\ (1998) -- thick solid crosses. The diagonal crosses represent lower limits 
to the background intensity inferred from source counts. From left to right: the 
850-$\mu$m count of Blain et al.\ (1999b); the 450-$\mu$m count of Smail et al.\ 
(1997), 
as updated to include more recent information from Ivison et al.\ (1999);
and the 175- and 95-$\mu$m counts of Kawara et al.\ (1998) and Puget et al.\ 
(1999). The 
background spectrum predicted in the 35-, 40-, 45- and 50-K models (Table\,1) 
are represented by solid, dashed, dot-dashed and dotted lines respectively, 
and are plotted across the entire abscissa. In a recent paper, Lagache et al.\ 
(1999) claim to have detected a warm diffuse Galactic dust component that 
accounts for about 50\,per cent of the isotropic DIRBE signal attributed 
to the extragalactic background intensity by 
Hauser et al.\ (1998) and Schlegel et al.\ (1998).  
}
\end{figure}

The bright low-redshift 60-$\mu$m count, 
\begin{equation} 
N_{60} \propto \sqrt{ { {x_0^3} \over {(F \sigma)_0 (1-f_{\rm A})^3} } } \,
S_{60}^{-3/2},
\end{equation}
is independent of the cosmological model. 
At $S_{60} = 10$\,Jy, $N_{60} = 19\pm2$\,sr$^{-1}$ (Saunders et al.\ 1990). 
A value of $\sqrt{ x_0^3 / (F \sigma)_0 (1-f_{\rm A})^3} = 
(11 \pm 2) \gamma \times 10^{-7}$\,Gyr$^{-1/2}$ provides a good fit for any dust 
temperature between 30 and 60\,K. The values of the normalization
of the luminosity density $x_0/(1-f_{\rm A})$ and the activity parameter 
$(F\sigma)_0^{-1}$ 
at $z \simeq 0$ that are required to fit the background spectrum and 
60-$\mu$m counts depend on the fluctuation index $\gamma$ as 
$\gamma^{-1}$ and $\gamma$ respectively. 
Thus the mass-to-light ratio of merging galaxies (equation 12) 
is expected to be independent of the value of the perturbation index. 
The values of $F\sigma$ listed in Table\,1 are lower limits to the time-scale of 
the starburst/AGN $\sigma$, as $F \le 1$. They are generally consistent 
with the starburst time-scales of order $5 \times 10^7$\,yr derived by 
Mihos \& Hernquist (1996) from hydrodynamical simulations of galaxy 
mergers.  

The background radiation spectra and 60-$\mu$m counts corresponding to the 
models listed in Table\,1 are shown in Figs\,3 and 4(a) respectively. Note 
that the inclusion of positive evolution in the efficiency parameter $x(z)$ 
overcomes the underprediction of the 60-$\mu$m counts in the 
constant-$x$ model of Blain \& Longair (1996); see their fig.\,6. 
The slope of the 60-$\mu$m counts in the new model also more adequately 
represents the data than the predictions of Guiderdoni et al. (1998). 
The luminosity function of nearby dusty galaxies at 60-$\mu$m (Saunders et al.\ 
1990; Soifer \& Neugebauer 1991) predicted in the 40-K model, which is 
calculated at $z=0$ by evaluating $F\sigma\dot N_{\rm form}$ at the mass  
corresponding to a luminosity $L$ in equation (12), is shown in Fig.\,4(b). The 
form of the luminosity function depends on the value of the
perturbation index $\gamma$, even though the 60-$\mu$m counts 
are independent of $\gamma$. The best representation of the observed function is 
provided by a value of $\gamma = 2/3$, or $n \simeq -1$. A similar
high-luminosity slope could be achieved by modifying the form of equation (12) 
so that $L \propto M^\beta$, where $\beta > 1$. However, to match the
observations with a scale-independent value of $n=1$, 
a rather extreme value of $\beta=2$ is required. In the 
work that follows $n$ is assumed to take the value $-1$, similar to the value
found for this range of masses by Peacock \& Dodds (1994). In this case, the 
faint-end slope of the low-redshift 60-$\mu$m luminosity function is equivalent 
to a Schechter function parameter $\alpha = -5/3$ (Schechter 1976). This is 
the same as the faint-end slope $\alpha = -1.60 \pm 0.13$ of the 
optically selected luminosity function derived by Steidel et al.\ (1999) at 
$z \simeq 3$, which describes the most numerous population of 
high-redshift galaxies that are actively forming stars. 
At $0.75 < z < 1$, the luminosity function of 
the blue star-forming 
galaxies in the CFRS survey also has a faint end slope $\alpha = -1.56$ 
(Lilly et al.\ 1995). 
Steep faint-end slopes with indices of about 
$-1.8$ are expected for the luminosity functions of dwarf, irregular and 
infrared-luminous galaxies, as discussed by Hogg \& Phinney (1997). 

\begin{figure*}
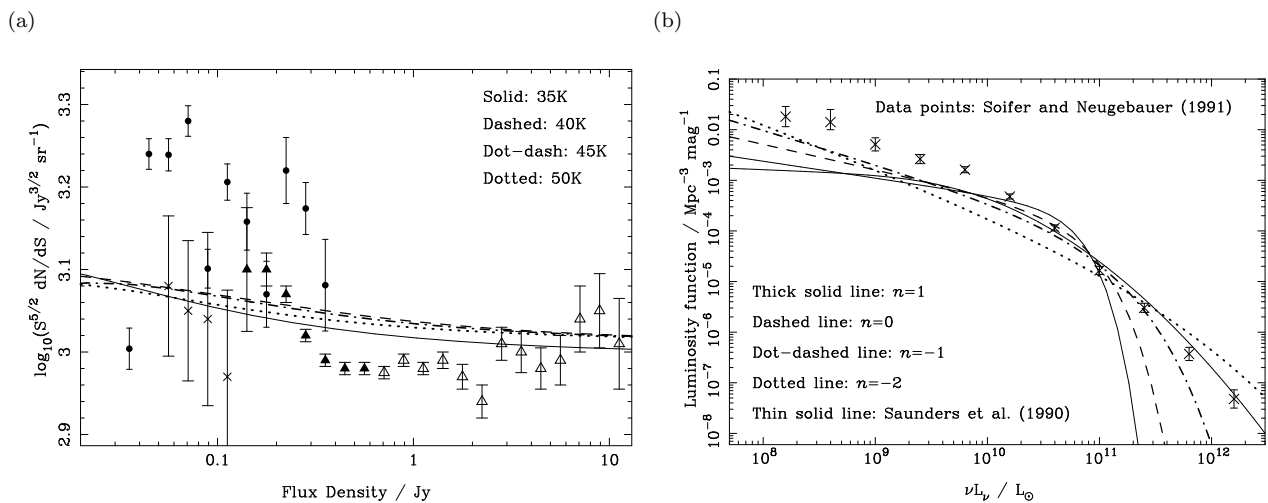

\begin{minipage}{170mm}
(a) \hskip 81mm (b)
\begin{center}
\epsfig{file=fig4a.ps, width=5.75cm, angle=-90} \hskip 5mm
\epsfig{file=fig4b.ps, width=5.75cm, angle=-90}
\end{center}
\caption{(a) Observed 60-$\mu$m counts of {\it IRAS} galaxies and the 
best-fitting models listed in Table\,1, plotted in the format used by Oliver,
Rowan-Robinson \& Saunders (1992). The data are taken from Hacking \& Houck 
(1987; crosses), Rowan-Robinson et al.\ (1990; empty triangles), Saunders et al.\ 
(1990; filled triangles) and Gregorich et al.\ (1995; circles): see also Bertin, 
Dennefeld \& Moshir  
(1997). (b) The 60-$\mu$m luminosity function predicted in the 40-K model, for 
four different values of the primordial fluctuation index $n$. The results are 
compared with the luminosity function of Saunders et al. (1990) and Soifer 
\& Neugebauer (1991). The luminosity functions derived for the other 
assumed dust temperatures are very similar.
}
\end{minipage}
\end{figure*}
 
\subsection{Self-consistency} 

In section\,4 of Blain et al.\ (1999c) the self-consistency of 
models of dust-obscured galaxy formation was discussed. 
We demanded that a sufficiently large mass of metals, and 
associated dust, was required to be generated by nucleosynthesis at each 
epoch to account for the far-infrared emission 
predicted by the model. As the 
mass of dust required to generate a given far-infrared luminosity depends 
strongly on the dust temperature, this consistency condition is most easily 
expressed as a minimum dust temperature at each redshift. In the 
cases of the models listed in Table\,1, this lower limit to the dust temperature 
is presented in Fig.\,5, both with and without an assumed high-redshift 
Population III to generate dust. If 2\,per cent of the total star formation activity 
takes place in a high-redshift Population III, then the self-consistency limits 
are always satisfied.  The details of the calculations can be found in Blain 
et al.\ (1999c). It is assumed that all the dust that is generated prior to a 
particular redshift still existed at that redshift and was available to absorb 
and reprocess the light from young hot stars and AGN. However, in an 
hierarchical model, only a small fraction of all the dust generated by any 
epoch is found in galaxies actively involved in a merger at that epoch; 
this fraction increases from about 8 to 25\,per
cent progressively from the 35-K to 50-K model listed in Table\,1. Thus, because 
most of the dust will be found in quiescent objects, the condition in 
Fig.\,5 is less severe than it might be. However, even if only 10\,per 
cent of all dust is involved in a luminous dust enshrouded merger, then 
the lower limit to the temperature increases only by about 50\,per
cent, and so this consistency requirement is not difficult to satisfy. The same 
correction is required if 90\,per cent of the energy generated in merging 
galaxies is attributable to accretion onto AGN, that is, if the AGN fraction 
$f_{\rm A}=0.9$. Hence, the models are readily self-consistent if high-redshift 
Population-III stars exist to generate early metals. More sophisticated models 
of the coupled evolution of dust, 
gas and stars have been presented by Pei \& Fall (1995), Eales \& Edmunds 
(1996) and Pei et al.\ (1999). 

\begin{figure}
\begin{center}
\epsfig{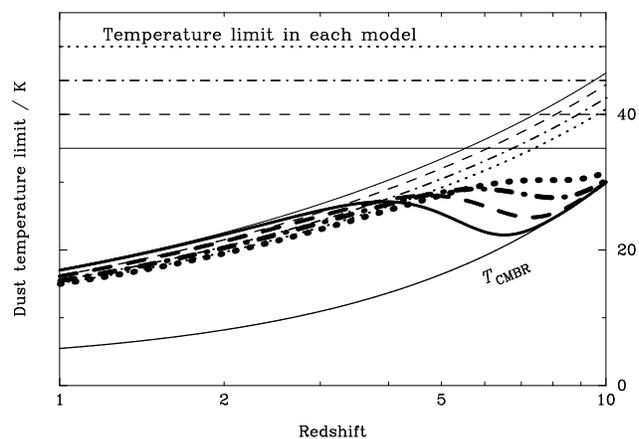}
\end{center}
\caption{The lower limits to the dust temperature as a function of 
redshift required by the consistency argument discussed in Section\,2.5.  
The dust temperatures in the 35, 40, 45, and 50-K models are shown by 
the horizontal lines, and the lower limit imposed by the CMBR temperature 
$T_{\rm CMBR}$ is shown by the lower solid curve. In a self-consistent model, 
the lower limit to the temperature must be less than the plotted temperature 
limits at all epochs. The line styles are the same as those in Figs\,3 and 4(a). 
The four thin lines show the lower limits if no Population III stars are included. 
The four thick lines show the lower limits if 2\,per cent of all star formation 
activity takes place in a high-redshift Population III. $z_0=20$ in all the models.}
\end{figure}

\begin{figure*}
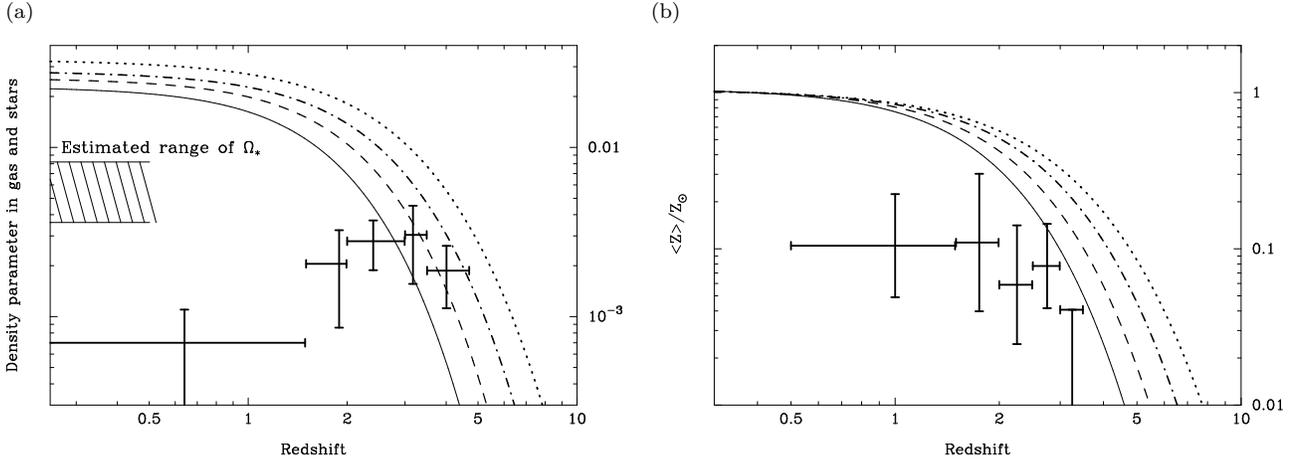

\begin{minipage}{170mm}
(a) \hskip 81mm (b)
\begin{center}
\vskip -2mm
\epsfig{file=fig6a.ps
, width=5.45cm, angle=-90} \hskip 5mm
\epsfig{file=fig6b.ps
, width=5.45cm, angle=-90}
\end{center}
\caption{(a) The density parameter of gas and stars as a function of redshift. The 
line styles represent the same models as those in Figs.\,3, 4(a) and 5. The
curves represent the density parameter of material processed into stars in the
models listed in Table\,1, assuming a Salpeter IMF with mass limits of
0.07 and 100\,M$_\odot$. The data for the total density of stars at the present
epoch (shaded region) was obtained by Gnedin \& Ostriker (1992). The data
points represent the density parameter in neutral hydrogen (Storrie-Lombardi
et~al.\ 1996). (b) The rate of increase of metallicity expected in the same models, 
with data from Pettini et al.\ (1997).
}
\end{minipage}
\end{figure*}

\subsection{Metal enrichment and the production of low-mass stars}

At the present epoch the density of metals formed by nucleosynthesis in stars 
$\Omega_{\rm m}(0)$ (equation 9) is expected to be about $10^{-3}$ in the 
models presented in Table\,1. If solar metallicity, about 2.5\,per cent by mass 
(Savage \& Sembach 1996), is typical of the Universe as a whole, and the 
density parameter in baryons $\Omega_{\rm b}h^2=0.019$ (Burles \& Tytler 1998), 
then $\Omega_{\rm m}\simeq 1.9\times10^{-3}$ if $h=0.5$. Thus all the models 
listed in Table\,1 are consistent with this limit, even if the AGN fraction 
$f_{\rm A}=0$ and all the luminosity of merging galaxies is due to star 
formation activity that generates heavy elements.

The density parameter in the form of stars at the present epoch 
$\Omega_*(0)=(5.9\pm2.3)\times10^{-3}$ (Gnedin \& Ostriker 1992). Observations 
of Lyman-$\alpha$ absorbers along the line of sight to distant quasars allow the 
evolution of the mass of neutral gas and the typical metallicity in the Universe 
to be traced as a function of epoch (Storrie-Lombardi, McMahon \& Irwin 
1996; Pettini et al.\ 1997). In Fig.\,6(a) the mass of material that has been 
processed into stars is derived as a function of epoch in each of the 
star-formation histories listed in Table\,1, assuming that the AGN fraction 
$f_{\rm A}=0$ and a Salpeter initial mass function (IMF) with a lower 
mass limit of 0.07\,M$_\odot$. In this case about 65--70\,per cent of all stars 
formed are still burning at the present epoch. The values of $\Omega_*(0)$ 
predicted are thus about 3 times larger than the observed value, 
but are comparable with the values derived 
in our earlier models (Blain et al.\ 1999c). In order to account for this difference, 
either a lower mass limit to the IMF of about 1\,M$_\odot$ or a value of the AGN 
fraction $f_{\rm A} \simeq 0.75$ is required. This high-mass IMF would be 
compatible with the inferred lower limit to the IMF of 3\,M$_\odot$ required by 
Zepf \& Silk (1996) to explain the mass-to-light ratios of elliptical galaxies, 
and by Rieke et al.\ (1993) to interpret observations of M82. Stars with 
masses less than 3\,M$_\odot$ appear to be less numerous than expected 
from a Salpeter IMF in recent observations of the low-redshift starburst 
galaxy R136 (Nota et al.\ 1998). Goldader et al.\ (1997) report that the 
results of near-infrared spectroscopy of nearby {\it IRAS} galaxies with 
luminosities between 10$^{11}$ and 10$^{12}$\,L$_\odot$ support a deficit 
of stars with masses less than 1\,M$_\odot$ in these systems. 
More details about variations in the 
high-redshift IMF are discussed by Larson (1998). 

Metals appear to be 
overproduced by about a factor of 5 at redshifts of 2 and 3 in the hierarchical 
models, as shown in Fig.\,6(b), but again these results can be reconciled with 
the observations if a significant fraction of the 
luminosity of dusty galaxies is being powered by accretion on to AGN. Note, 
however, that the observations of metallicity could be biased against metal-rich 
regions of the Universe, either because of their small physical size 
(Ferguson, Gallagher \& Wyse 1998) or because of the complete obscuration of 
a fraction of background QSOs (Fall \& Pei 1993). {\it ASCA} X-ray 
observations of 
significant enrichment in intracluster gas (Mushotzky \& Loewenstein 1997; 
Gibson, Loewenstein \& Mushotzky 1998), could indicate that this is the case, 
at least in high-density environments.

The observed turn-over in the neutral gas fraction and the maximum rate of star 
formation shown in Fig.\,6(a) are approximately 
coincident in redshift, and the rate of enrichment in the hierarchical models is 
broadly consistent with the slope interpolated between the three highest redshift 
data points plotted in Fig.\,6(b).

\begin{figure}
\begin{center}
\epsfig{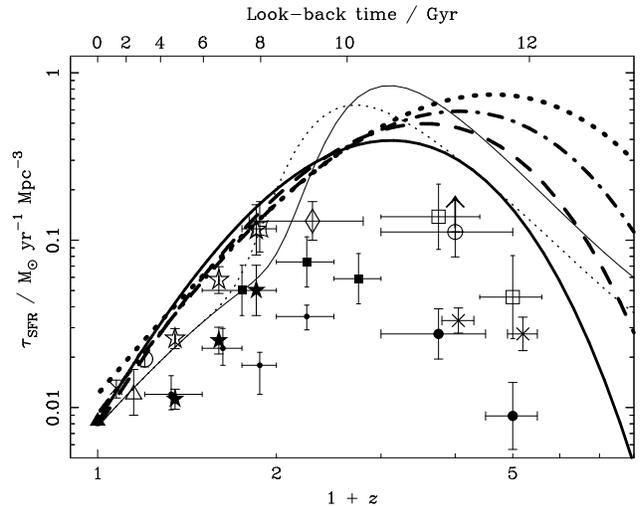} \hskip 7mm
\end{center}
\caption{The luminosity density predicted by the 35-, 40-, 45- and 50-K 
models (thick solid, dashed, dot-dashed and dotted lines respectively) listed 
in Table\,1. The values are converted into a star formation rate assuming the
same conversion rate as Blain et al. (1999c). References to the points on the 
plot are listed in the caption of Fig.\,1. The thin solid and dotted lines trace 
the star formation histories derived in Blain et al.\ (1999c) and 
Barger et al.\ 
(1999b), which were shown in Fig.\,1.
}
\end{figure}

\subsection{The history of star-formation/AGN fuelling} 

In Fig.\,7 the form of evolution of the luminosity density is shown 
as a function of redshift in the models listed in Table\,1. All the 
models predict curves with a rather similar form at $z \ls 2$. The transformation 
between luminosity and star formation rate is the same as that assumed by 
Blain et al.\ (1999c): that is, a SFR of 1\,M$_\odot$\,yr$^{-1}$ is 
equivalent to a luminosity of $2.2 \times 10^9$\,L$_\odot$. At low redshifts the 
evolution of luminosity density is consistent with optical and near-infrared 
observations, and with the results presented in our earlier paper.

\begin{figure*}
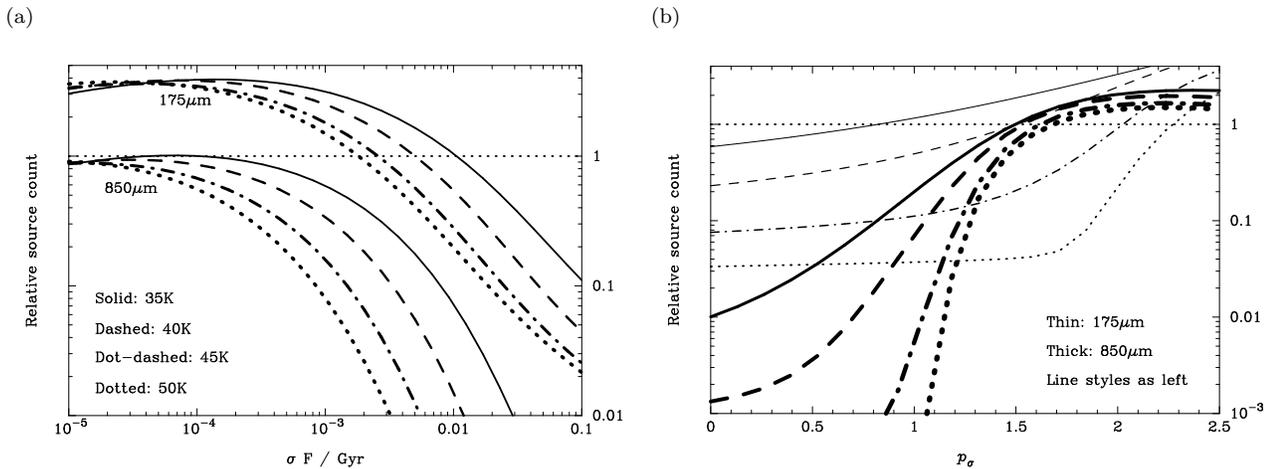

\begin{minipage}{170mm}
(a) \hskip 81mm (b)
\begin{center}
\epsfig{file=fig8a.ps, width=5.35cm, angle=-90} \hskip 5mm
\epsfig{file=fig8b.ps, width=5.35cm, angle=-90}
\end{center}
\caption{The ratio of the counts predicted by the four models listed in Table\,1, 
and the observed counts, at 175 and 850\,$\mu$m (Kawara et al. 1998; 
Blain et al. 1999b respectively). The upper limits to the 
counts at 450\,$\mu$m and 2.8\,mm are never violated. In (a) the ratio is
shown as a function of the reciprocal of the activity parameter $F\sigma$, 
which is fixed as a function of redshift; 
in (b) the ratio is shown as a 
function of $p_\sigma$, the exponent in equation (18), which describes the 
exponential evolution of the activity parameter 
$(F\sigma)^{-1} \propto (F\sigma)_0^{-1} \exp{p_\sigma z}$. 
Clearly, strong redshift 
evolution of the activity parameter in 
merging galaxies is required in order to fit the counts at long 
wavelengths.}
\end{minipage}
\end{figure*}

\section{Source counts}

\subsection{Fitting the available data} 

The models presented in Table\,1 were constrained using the properties of the 
counts of the low-redshift 60-$\mu$m {\it IRAS} galaxies. The same formalism 
can be used to determine the counts of more distant dusty galaxies in the 
mid-/far-infrared and millimetre/submillimetre wavebands, where a large amount 
of additional information about the surface density of more distant dusty galaxies 
is available. There is an upper limit to the surface density of sources at 2.8\,mm 
(Wilner \& Wright 1997); counts at 850\,$\mu$m (Smail et al.\ 1997; Barger et al.\ 
1998; Holland et al.\ 1998; Hughes et al.\ 1998; Barger et al.\ 1999a; 
Blain et al.\, 1999b; Eales et al.\ 1999); upper limits (Smail et al.\ 1997; 
Barger et al.\ 1998), and a new count (Blain et al.\ 2000) at 450\,$\mu$m; 
175-$\mu$m {\it ISO} counts from Kawara et al.\ (1998) and Puget et al.\ (1999); 
95-$\mu$m counts from Kawara et al.\ (1998); and 7- and 15-$\mu$m counts 
from an extremely deep {\it ISO} image of Abell\,2390 (Altieri et al.\ 1999), 
which yields counts that are even deeper than those determined in blank-field 
surveys by Oliver et al.\ (1997), Aussel et al. (1999) and Flores et al.\ (1999).

If the values of the activity parameter at redshift zero, $(F\sigma)_0^{-1}$, 
listed in Table\,1 are used to estimate the counts of galaxies at 850 and 
175\,$\mu$m, then the results 
underpredict the observed counts by a large factor. The form of evolution of the
merger efficiency $x(z)$ is fixed by the observed background radiation intensity, 
and so, keeping within the framework of our well-constrained models, the value 
of the activity parameter $(F\sigma)^{-1}$ at high redshift must be allowed to 
increase above its value at redshift zero in order to account for the 
observations. This has the effect of increasing the luminosity of 
high-redshift mergers, thus increasing the 175- and 850-$\mu$m counts. 
However, the background radiation intensity and the low-redshift 60-$\mu$m 
counts remain unchanged. 

The form of evolution of the activity parameter $(F\sigma)^{-1}$ 
that is required to explain the 
data is illustrated in Fig.\,8. In Fig.\,8(a) the ratio of the model predictions and 
the observed counts at wavelengths of 175 and 850\,$\mu$m (Kawara et al. 
1998; Blain et al.\ 1999b respectively) are compared as a function of 
the activity parameter in the four models listed in Table\,1. The same value of 
the activity parameter cannot account for the observed counts at both 
wavelengths simultaneously, and the value required to explain the 
low-redshift 60-$\mu$m counts is different from either. 
The value of the activity parameter required to fit the 60-, 175- and 
850-$\mu$m counts increases monotonically. Because the median redshift of 
the galaxies contributing to the counts at these redshifts is expected to increase 
monotonically, in Fig.\,8(b) we present the ratio of the model predictions and 
the observed counts as a function of a parameter $p_\sigma$ that describes 
a simple form of exponential redshift evolution of the activity parameter,
\begin{equation} 
(F\sigma)^{-1} = (F\sigma)^{-1}_0 \exp{ p_\sigma z}.  
\end{equation} 
The exponential form provides a reasonable fit to the data, but is only one 
example of a whole family of potential functions. The important feature is that 
the function chosen to represent the activity parameter $(F\sigma)^{-1}$ 
increases rapidly with increasing redshift. 
 
The zero-redshift value of the activity parameter $(F\sigma)^{-1}_0$ is fixed by 
requiring that the low-redshift 60-$\mu$m count prediction is in agreement with 
observations; see Table\,1. The values of the evolution parameter 
$p_\sigma$ that correspond to the most reasonable fit for assumed single dust 
temperatures of 35, 40, 45 and 50\,K are about 1.5, 1.5, 2.0 and 2.3 respectively. If 
the specific form of the redshift evolution of the activity parameter
$(F\sigma)^{-1}$ shown in equation (18) is assumed, 
then a dust temperature of 35 or 40\,K is most 
consistent with the data, the same temperature that was required for 
consistency by both Blain et al.\ (1999c) and Trentham et al.\ (1999), and 
is in agreement with the dust temperatures derived for high-redshift 
QSOs by Benford et~al.\ (1999).

The increase in the value of the activity parameter $(F\sigma)^{-1}$ 
as a function of 
redshift can be interpreted in terms of two extreme scenarios, or as a 
combination of both. In the first scenario, the fraction of dark halo mergers that 
lead to a luminous phase in a dusty galaxy $F$ is fixed, but that the duration of 
the luminous phase $\sigma$ is less at high redshifts. This is plausible, based 
on the results of simulations of galaxy mergers (e.g. Mihos 1999; Bekki
et al.\ 1999); on 
average, the typical mass of a merging pair of galaxies is expected to be less at 
high redshifts in an hierarchical scenario of galaxy evolution, and the gas content 
of the galaxies is expected to be greater. As a result, the dynamical time of a
merger would be expected to decrease with increasing redshift, and the 
viscosity of the ISM would be expected to increase. Both of these factors might 
be expected to increase the star formation efficiency of a merger with 
increasing redshift. 
In the second scenario, the duration of the luminous phase associated
with a merger $\sigma$ is independent of redshift, but the fraction of 
mergers that induce such a phase $F$ is reduced as redshift increases. It 
is perhaps more plausible that the second of these scenarios could produce the 
large change in the activity parameter $(F\sigma)^{-1}$, by a factor of about 100
from $z=0$ to $z=3$ that is required to fit the data. This is 
because the duration of the luminous phase of a merger-induced starburst 
$\sigma$ must exceed the lifetime of a reasonably massive star, i.e. 
$\sigma \gs 10^7$\,yr. If star formation activity powers a significant fraction of 
the SCUBA galaxies, as seems reasonable, then this limit to the value of the 
merger duration $\sigma$ is constrained to be greater than about 10$^7$\,yr, 
only a few times less than the values of $\sigma$ at $z=0$ listed in Table\,1. 
Thus it seems likely that a large fraction of the increase in the value of the 
activity parameter $(F\sigma)^{-1}$, 
which is required at high redshifts to account for the observed counts, 
should be attributed to a reduction in the fraction $F$ of dark halo  
mergers that generate a luminous galaxy. We speculate that this may be 
connected with the lower typical metallicity expected at higher redshifts. In 
a lower metallicity system the cooling of dense gas would be expected to 
be less efficient, and so a large amount of 
high-mass star formation may be unable to 
take place during the short merger process.

\subsection{Predicted counts of dusty galaxies} 

Counts predicted by the four models listed in Table\,1, employing the 
values of $p_\sigma$ listed above, are compared with observations at 
wavelengths of 15, 60, 175, 450, 850, 1300 and 2800\,$\mu$m in Fig.\,9. While these 
models do not present unique solutions, fewer parameters are involved in the 
model than the number of separate pieces of constraining data. In future, by 
comparing the 
predictions of the models with observations, especially with the redshift 
distributions of the SCUBA galaxies (Barger et al.\ 1999b; Lilly
et al.\ 1999; Smail et al.\ 1999, in preparation), the models can be developed 
to account more accurately for the increasing amount of available data.

\begin{figure*}
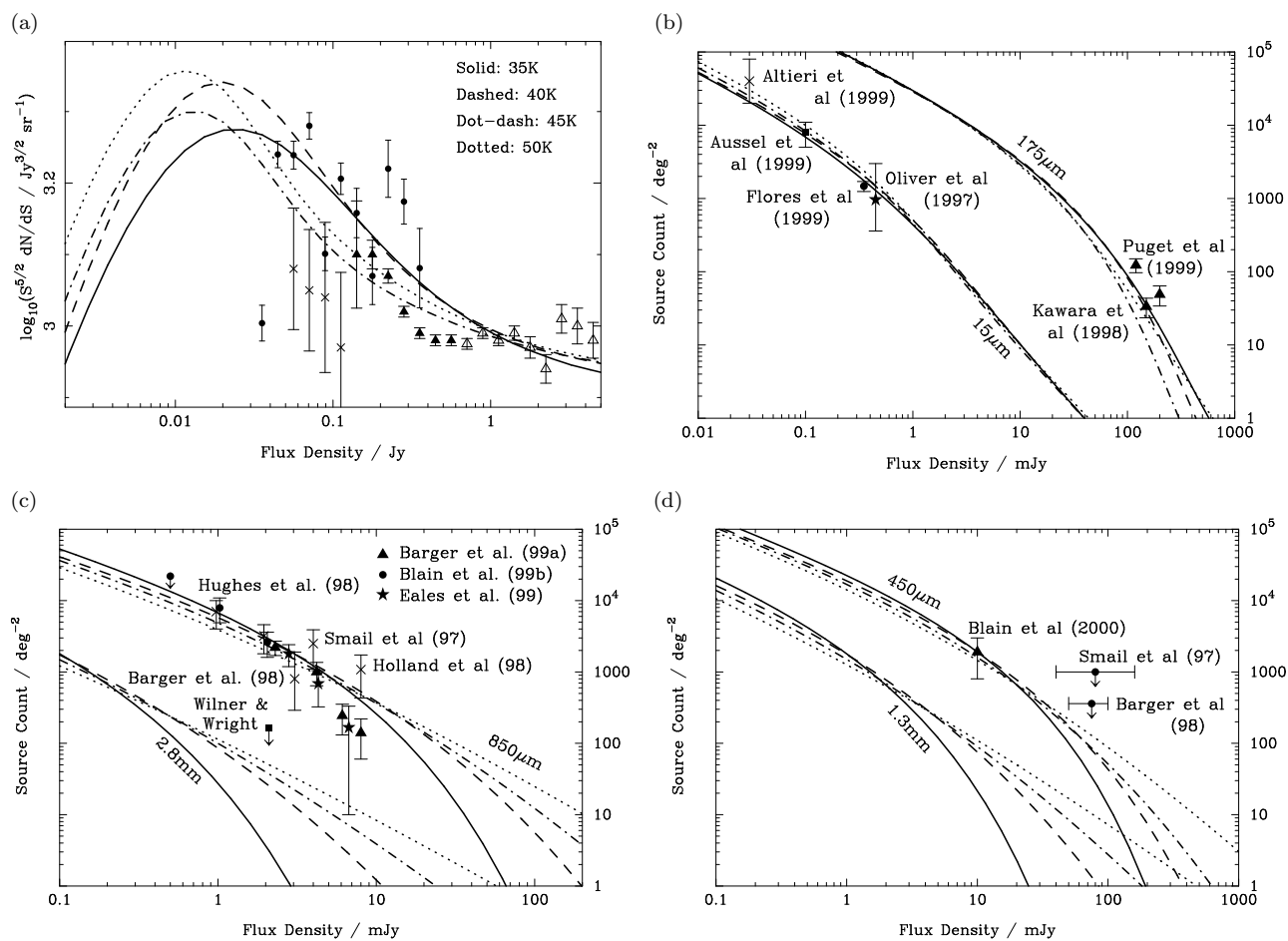

\begin{minipage}{170mm}
(a) \hskip 81mm (b)
\begin{center}
\vskip-4mm
\epsfig{file=fig9a.ps, width=5.6cm, angle=-90} \hskip 5mm %\vskip -1mm
\epsfig{file=fig9b.ps, width=5.75cm, angle=-90}
\end{center}
(c) \hskip 81mm (d)
\begin{center}
\vskip -4mm
\epsfig{file=fig9c.ps, width=5.6cm, angle=-90} \hskip 4mm
\epsfig{file=fig9d.ps, width=5.6cm, angle=-90}
\end{center}
\caption{Counts predicted by the models listed in Table\,1, compared with
available data. The 60-$\mu$m counts are shown in (a), those at 15 and
175\,$\mu$m are shown in (b), those at 850\,$\mu$m and 2.8\,mm are shown in
(c), and those at 450\,$\mu$m and 1.3\,mm are shown in (d). The
references to the data in (a) are given in the caption of Fig.\,4(a); in 
(b), (c) and (d) they are written adjacent to the data points.
}
\end{minipage}
\end{figure*}

\subsection{The corresponding radio counts} 

There is a tight correlation between the flux densities of low-redshift 
galaxies in the radio and far-infrared wavebands (see the review by Condon 
1992). Thus the counts of faint galaxies observed in the 
radio waveband should not be 
overproduced when the SEDs of the galaxies in the 35-, 40-, 45- and 50-K 
models presented here are extended into the radio waveband using this 
correlation. It is
permissible to underpredict the counts, as there will be a contribution from 
AGN to the faint counts, which need not be associated with powerful restframe 
far-infrared emission from dust. Partridge et al.\ (1997) report a 
8.4-GHz galaxy count of $1.0 \pm 0.1$\,arcmin$^{-2}$ brighter than a flux 
density of 10\,$\mu$Jy. The correspondings count predicted by the 35-, 40-, 45- 
and 50-K models are 0.6, 0.7, 1.0 and 0.6\,arcmin$^{-2}$ respectively. Thus 
all the models 
discussed here are consistent with the deep 
radio observations. For comparison, a
count of 0.8\,arcmin$^{-2}$ is predicted by the modified Gaussian model 
discussed in Barger et al.\ (1999b), which is modified from the results of the 
simple luminosity evolution models presented by Blain et al.\ (1999c). 

\subsection{Redshift distributions}

The redshift distributions of submillimetre-selected sources at, or just below, the 
flux density limits of current surveys have been discussed by Blain et al.\ (1999c)
in the context of models of a strongly evolving population of distant dusty 
galaxies, based on the 
low-redshift {\it IRAS} galaxy luminosity function. The 
first spectroscopic observations of a large fraction of the potential optical 
counterparts to SCUBA galaxies identified in deep multicolour optical images 
(Smail et al.\ 1998) have been made by Barger et al.\ (1999b) (see Fig.\,10). This 
redshift distribution is consistent with the optical 
identifications made by Lilly et al.\ 
(1999) in a SCUBA survey of Canada--France Redshift Survey (CFRS) fields. The 
distribution shown in Fig.\,10 is, however, subject to potential misidentifications 
of SCUBA galaxies. For example, recent deep near-infrared images 
show that two of the Smail et al.\ (1998) SCUBA galaxies, which were originally 
identified with low-redshift spiral galaxies, can more plausibly be associated with 
extremely red objects (EROs) that were unidentified in optical images (Smail et
al.\ 1999). In two other cases, at $z=2.55$ (Ivison et al.\ 1999) and $z=2.81$ 
(Ivison et al.\ 1998), the identifications have been confirmed by 
detections of redshifted CO emission (Frayer et al.\ 1998; 1999), and in another 
case spectroscopy and {\it ISO} observations (Soucail et al.\ 1999) 
strongly support the identification of a ring galaxy at $z=1.06$.

The preliminary redshift distribution of SCUBA galaxies, shown in Fig.\,10, is 
broadly consistent with the predictions of the Gaussian model of Blain et al.\ 
(1999c). A modified Gaussian model, as shown in Fig.\,1, was described by  
Barger et al.\ (1999b); the values of the evolution parameters in the modified 
Gaussian model were explicitly fitted both to the background radiation intensity 
and count data and to the observed median redshift. In Fig.\,10 the observed 
redshift distribution is compared with the redshift distributions predicted in the 
Gaussian and modified Gaussian models, and with the predictions of the 
hierarchical models developed here (see Table\,1). Median redshifts of about 
2.2, 2.7, 3.2 and 3.5 are expected in the 35-, 40-, 45- and 50-K 
hierarchical models respectively. 

The redshift distributions predicted by the hierarchical models have median 
redshifts greater than that in the modified Gaussian model, but less than 
those in either the other models presented by Blain et al.\ (1999c) or the 
hierarchical model E from Guiderdoni et al.\ (1998), all of which provide a 
reasonable fit to both the background radiation intensity 
and source counts in the far-infrared/submillimetre waveband. Based on these 
results, the coolest 35-K model seems to be in best agreement with the 
available data. A model in which the single-temperature dust clouds discussed 
here are replaced by a temperature distribution will probably be required to 
account for the redshift distribution of the SCUBA galaxies. When the two spiral 
galaxies at $z < 0.5$ are replaced by EROs at $z>1$ (Smail et al.\ 1999), 
the agreement between the 35-K hierarchical prediction and the observed 
redshift distribution is rather satisfactory. 

\begin{figure}
\begin{center}
\epsfig{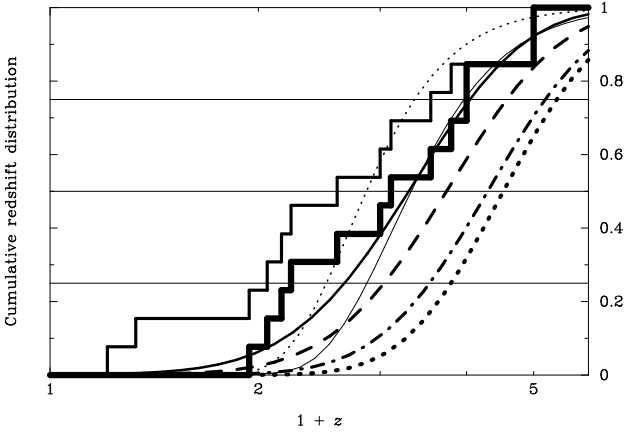} %\hskip 7mm
\end{center}
\caption{The redshift distributions of 850-$\mu$m galaxies in the Smail 
et al.\ (1998) sample predicted in the 35-, 40-, 45- and 50-K hierarchical models 
listed in Table\,1 (thick smooth lines; solid, dashed, dot-dashed and dotted
respectively), the Gaussian model (thin smooth 
solid line; Blain et al.\ 1999c) and the modified Gaussian model (thin dotted line; 
Barger et al.\ 1999). The modified Gaussian model is fitted to the median 
redshift of the preliminary redshift distribution determined by Barger et al.\ 
(1999b) for the optical counterparts to submillimetre-selected galaxies (Smail et
al.\ 1998), which is shown by the thinner jagged solid line. Sources with 
no optical counterparts are assumed to be at $z=4$. Using subsequent 
$K$-band observations, Smail et al.\ (1999) have modified the identifications of 
the counterparts of the two galaxies at $z<1$ in the sample, from low-redshift 
spirals to high-redshift EROs. After this modification, the revised redshift 
distribution is shown by the thicker jagged solid line, if the EROs are both 
assumed to lie at $z=3$.}
\end{figure}

In all the hierarchical models, despite strong negative evolution of the 
mass-to-light ratio of mergers with increasing redshift, most of the 
detected galaxies are expected to lie at redshifts less than 5, and so will be 
accessible to multi-waveband study using 8-m class telescopes. When final 
reliable identifications and redshifts for submillimetre-selected galaxies 
are available, this information will be crucial for refining the hierarchical 
model. 

\section{Optical backgrounds and counterparts} 

The discussion has so far centred on the properties of merging galaxies as 
observed through their dust emission in the mid-infrared, far-infrared and 
millimetre/submillimetre wavebands. Here we assume the same forms of 
evolution of both the merger efficiency parameter $x$ and the activity 
parameter $(F\sigma)^{-1}$ that were required to account for the data
in the far-infrared and submillimetre wavebands in the previous section, but 
make predictions in the near-infrared, optical and ultraviolet wavebands. In 
particular, we investigate the 35-K model, in which the redshift distribution 
of SCUBA galaxies is in best agreement with observations. 

Subject to the uncertain fraction of the luminosity of these galaxies that 
is assumed to be powered by star formation activity, we predict the integrated 
background radiation intensity from the near-infrared to ionizing 
ultraviolet wavebands, and the counts of galaxies with SEDs that are 
dominated by evolved stars in the near-infrared $K$-band, and by young stars 
in the optical $B$-band. By requiring that the $K$- and $B$-band 
counts are reproduced accurately, we estimate both the fraction of 
all energy released in mergers that is reprocessed by dust $A$ and the 
normalization of the activity parameter at $z=0$, $(F\sigma)_0^{-1}$ in the 
optical waveband. For a discussion of the evolution of faint galaxies and their 
stellar populations see Ellis (1997). 

\subsection{$K$-band counts} 

The counts of galaxies in the $K$-band at a flux density $S_K$ can be predicted 
by assuming the forms of the merger efficiency $x(z)$, as listed in Table\,1, an 
SED typical of evolved stars $f^K_\nu$ (Charlot, Worthey \& Bressan 1996), the 
Press--Schechter mass function (equation 1) and the mass-to-light ratio of 
evolved stellar populations $R_{\rm ML}$. The SED was calculated using a 
9.25-Gyr old Bruzual--Charlot instantaneous burst model with a Salpeter IMF. 
Upper and lower mass limits of 0.1 and 125\,M$_\odot$ were assumed for 
the IMF. Note that the form 
of the evolved stellar spectrum derived is almost independent of the exact 
values of the upper and lower mass limits assumed. The $K$-band count
\begin{equation} 
N_K (S_K) = \int_0^{z_0} \int_{M_K(z)}^\infty N_{\rm PS}(M)\,{\rm d}M\,
D(z)^2 \, { {{\rm d}r} \over { {\rm d}z} }\, {\rm d}z, 
\end{equation}
with 
\begin{equation} 
M_K(z) = {4 \pi D^2 (1+z) S_K} R_{\rm ML}(z) 
{ {\int f^K_{\nu'} \, {\rm d}\nu'} \over {f^K_{\nu_K (1+z)}} }.
\end{equation}
By ensuring that the predicted counts match the observed $K$-band counts, 
a suitable form of the mass-to-light ratio $R_{\rm ML}$ is determined as a 
function of redshift. The mass in this ratio is defined as the mass of the dark 
matter haloes of galaxies, taken from the Press--Schechter function 
(equation 1), and the luminosity is the bolometric luminosity of the evolved 
stellar population in the galaxies. 

\begin{figure*}
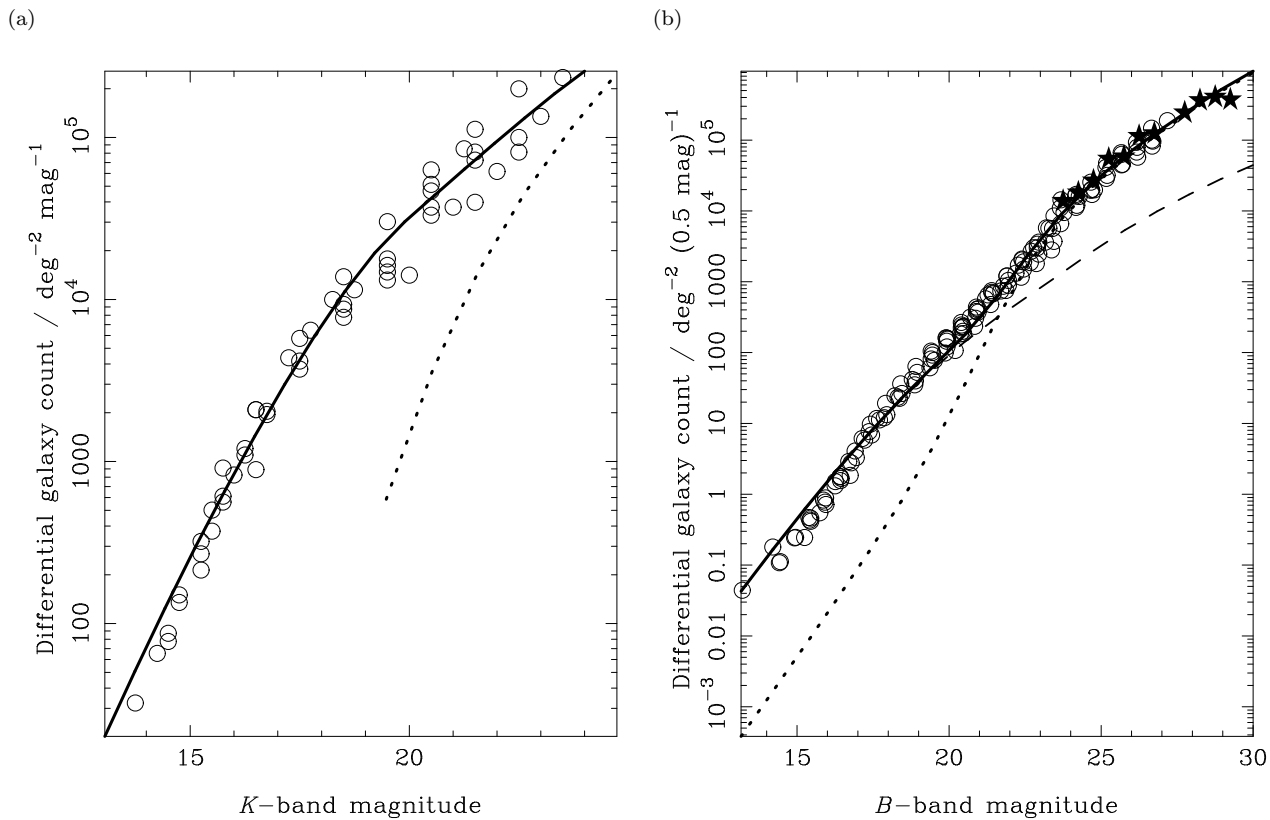

\begin{minipage}{170mm}
(a) \hskip 81mm (b)
\begin{center}
\epsfig{file=fig11a.ps, width=10.0cm, angle=-90} \hskip 5mm
\epsfig{file=fig11b.ps, width=10.0cm, angle=-90}
\end{center}
\caption{(a) Near-infrared $K$-band and (b) optical $B$-band counts predicted 
in the 35-K hierarchical model (Table\,1). The $K$-band counts are produced 
almost entirely by quiescent non-merging galaxies (solid line). 
If the flat spectra of the merging galaxies are extrapolated into the $K$ band, 
as shown by the dotted line, then the counts would be expected 
to increase by about 10\,per cent at $K=20.5$ and by 30\,per cent at $K=23$.  
In the $B$-band, the total counts 
(solid line) are dominated at the bright end by the small hot stellar component 
of the quiescent galaxies (dashed line), and at the faint end by the young stellar 
populations of flat-spectrum merger induced starbursts/AGN (dotted line). 
Data points obtained using ground-based telescopes are shown by open 
circles; data points from the {\it Hubble Space Telescope} are shown by 
filled stars. The data are taken from the compilation of Metcalfe et al. (1996).} 
\end{minipage}
\end{figure*}

\begin{figure*}
\begin{minipage}{170mm}
\begin{center} 
\epsfig{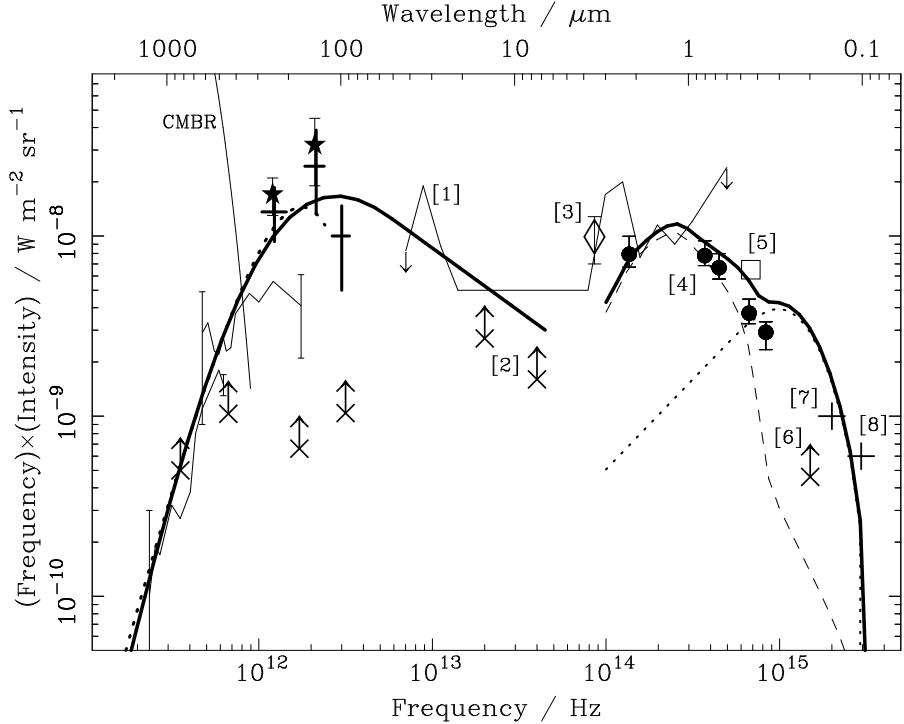}
\end{center}
\caption{The background radiation intensity in the hierarchical models from 
the millimetre to the ultraviolet waveband. The thick solid line shows the 
total background radiation intensity predicted in the 35-K model (Table\,1). In the
near-infrared/optical/ultraviolet waveband the separate contributions to the 
background intensity from the old stellar population (dashed line) and 
the young stellar population/AGN (dotted line) are also 
shown. At long wavelengths 
the data points are identical to those plotted in Fig.\,3. Other measurements are 
numbered: 1. Stanev \& Franceschini (1998); 2. Altieri et al. (1999); 
3. Dwek \& Arendt (1998); 4. Pozzetti et al. (1998); 5. Toller, Tanabe \& Weinberg 
(1987); 6. Armand, Milliard \& Deharveng (1994); 7. Lampton, Bowyer \& 
Deharveng (1990); and 8. Murthy et al.\ (1999); see 
also Bernstein, Freedman \& Madore (1999), who determine a greater absolute 
background radiation intensity in the optical waveband.} 
\end{minipage}
\end{figure*}

In order to reproduce the observed $B$-band counts, the counts derived for the 
evolved and merging components are added together, as shown in Fig.\,11(b).
The redshift dependence of the mass-to-light ratio is the same as that of the 
luminosity density of evolved stars, 
\begin{equation} 
\epsilon_{\rm L}(z) \propto { 1 \over {1+z} } 
{ { \int_0^{z_0} { {x(z')} \over {1+z'} } {\rm d}z' }  \over 
{ \int_z^{z_0} { {x(z')} \over {1+z'} } {\rm d}z' } },   
\end{equation}
which depends on the SFR at all earlier epochs. A factor of $1+z$ is included 
in the denominator to mimic the effects of passive stellar evolution (Longair
1998). At $z=0$ a form of the mass-to-light ratio, 
\begin{equation}
{ \displaystyle { { R_{\rm ML} } \over {{\rm M}_\odot {\rm L}_\odot^{-1}} } } 
= \left\{ \begin{array}{ll}
80, & {\rm if} \>\> L_{10} \ge 2;\\
117 L_{10}^{-0.55}, & {\rm otherwise}, 
\end{array}\right.
\end{equation}
where $L_{10} = L / 10^{10}$\,L$_\odot$, is required to match 
the observed $K$-band counts (Fig.\,11a) and the 
faint-end slope of the observed $K$-band luminosity function 
(Gardner et al.\ 1996; Szokoly et al.\ 1998). 

The well fitting $K$-band count that is derived from the model with this form of 
the mass-to-light ratio $R_{\rm ML}$ is shown, along with the observational 
data, in Fig.\,11(a).

\subsection{$B$-band counts} 

Both passive evolved galaxies and luminous merging galaxies make a 
contribution to the $B$-band counts. The evolved contribution is predicted by 
evaluating the function that produces the $K$-band count at the frequency of 
the $B$-band. The extrapolation is made using the model SED described above. 
An additional population of merging galaxies is also included. Their counts 
are determined using equation (13) directly, with a value of
\begin{equation} 
M_{\rm min} = { 1 \over {1-A}}
{ {4 \pi D^2 (1+z) S_B } \over { 0.007c^2 } } 
{ {F \sigma} \over { x(z) } } (1-f_{\rm A}) 
{ {\int f^B_{\nu'}\,{\rm d}\nu'} \over {f^B_{\nu_B(1+z)}} }.
\end{equation}
$A$ is the fraction of the total energy released in a merger that is reprocessed 
into the far-infrared waveband, and $f^B_\nu$ is the SED of a flat star-forming 
young stellar spectrum at frequencies less than the Lyman limit frequency 
$\nu_{\rm Ly} = 3.3 \times 10^{15}$\,Hz. 
$f^B_\nu \propto \nu^0$ if $\nu \le \nu_{\rm Ly}$ and zero otherwise. The blue 
power-law SED expected from an AGN will be described reasonably well by 
this SED at $\nu < \nu_{\rm Ly}$. 

The faint counts at $B > 21$, which are dominated by merging galaxies, can 
be reproduced in the model only if the dust absorption fraction 
$A \simeq 0.8$ and the zero-redshift 
activity parameter $(F\sigma)_0^{-1} = 2.5$\,Gyr$^{-1}$. The activity 
parameter incorporated in the model evolves with redshift as shown in 
equation (18), with the value of $p_\sigma = 1.5$ that is appropriate in 
the 35-K model. 
Note that this value of the activity 
parameter is less than that required to account for the observed 
submillimetre-wave counts, and that the ratio of energy emitted in the 
restframe ultraviolet and far-infrared wavebands is 1:4. This ratio is 
equivalent to 1.75 magnitudes of extinction when integrated over the optical 
and ultraviolet wavebands. 

The redshift distribution of faint galaxies at $z \gs 1$ derived in the 
hierarchical model is in good agreement with that observed for galaxies with 
$B<24$ (Cowie, Hu \& Songaila 1995); these details are discussed more 
extensively elsewhere (Jameson et al.\ 1999, in preparation). Note that the 
dependence of the submillimetre and faint $B$-band counts on the
merging efficiency parameter $x(z)$ and the AGN fraction $f_{\rm A}$ is identical, 
and so the value of the AGN fraction does not affect the resulting counts.

The most numerous population of faint high-redshift optically selected 
galaxies known are the Lyman-break galaxies (Steidel et al.\ 1996a, 1999) at 
$2.5 \ge z \ge 4.5$, which have apparent SFRs of a few 10\,M$_\odot$\,yr$^{-1}$. 
The surface density of detected Lyman-break galaxies is about an order of 
magnitude greater than that of submillimetre-selected galaxies, while their 
SFRs are typically about an order of magnitude less. 
 
The violence of a typical merger-induced starburst/AGN, and thus its 
detectability, is determined by the product of the merging efficiency 
and activity parameters, $x (F\sigma)^{-1}$, in the hierarchical model. The 
value of this composite parameter that is required to fit the observed 
submillimetre-wave counts is about 40 times greater than that required to fit the 
faint $B$-band counts. The value of the activity parameter $(F\sigma)^{-1}$ 
that is required to fit the submillimetre-wave counts is a factor of 
about 10 times 
greater than that required to fit the $B$-band counts. These differences are 
thus comparable with the observed ratios of the surface densities and 
luminosities of typical galaxies in the submillimetre-selected and 
Lyman-break samples. 

Based on these differences, we suggest a scenario in which the 
optically selected Lyman-break galaxies and the submillimetre-selected 
SCUBA galaxies are drawn from the same underlying population of 
luminous galaxy merger events, but are distinguished by being observed 
during two distinct phases of the merger process. We associate one 
phase, which is very luminous, short-lived and heavily dust enshrouded, with 
the SCUBA galaxies, and the other, which is less luminous and relatively 
lightly obscured, with the Lyman-break galaxies. During the first phase, which 
is about 40 times more luminous, but only about a tenth of the duration 
as compared with the second,  
most of the activity in the merger will be almost completely obscured from 
view in the optical waveband, but extremely bright in the submillimetre 
waveband. This phase is consistent with the extremely compact 
nuclear starburst/AGN activity observed by Downes \& Solomon (1998) 
on sub-kpc scales in nearby ultraluminous {\it IRAS} galaxies. During the 
second phase, less intense star formation activity would probably 
be distributed throughout the ISM of both merging galaxies. A short lived 
ultraluminous phase and a longer-lived less intense burst of
star formation activity during a merger are consistent with the results of 
hydrodynamic models of galaxy mergers by Mihos \& Hernquist (1996) and 
Bekki et al.\ (1999).

However, while plausible, this scenario is not necessarily correct. The 
faint counts in the submillimetre and optical waveband could simply 
be drawn from two distinct populations. The questions of whether and 
how ultraluminous dust-enshrouded mergers are connected with the 
Lyman-break galaxies can only be answered by making multiwaveband 
observations of large samples of submillimetre-selected galaxies in order 
to observe a time sequence of merging galaxies and to investigate the 
merger process in detail. Observations of the Lyman-break population in 
the submillimetre waveband (Chapman et al.\ 1999) will also help 
to address these questions.

\subsection{Integrated background light} 

The global luminosity density predicted by the hierarchical models is based on 
the evolution of the merging efficiency parameter $x(z)$. By making minor 
modifications to the formalism presented in Section 2.3, the models listed in 
Table\,1 can be used to predict the background radiation intensity. 

The near-infrared/optical/ultraviolet background radiation intensity produced by 
merging galaxies can be calculated as shown in equation (11), if the fraction of 
energy generated by a merger that is absorbed by dust $A$ is included. Thus 
\begin{equation} 
I^{\rm opt}_\nu \propto (1-A)
{\phi \over {\gamma \sqrt{\alpha}} } \int_0^{z_0} { {x(z)} \over {1-f_{\rm A}} }  
{ {\dot \delta(z)} \over {\delta(z)}} 
{ {f^B_{\nu(1+z)}} \over {\int f^B_{\nu'}\,{\rm d}\nu'} } 
\, { {{\rm d}r} \over {{\rm d}z}} \, { {{\rm d}z} \over {1+z} }. 
\end{equation}
The background radiation intensity due to the evolved stellar population is not 
given directly by the expression for the far-infrared background in equation (11). 
In that case the volume emissivity is given by equations (21) and (22). 
By integrating this emissivity of evolved objects over redshift, assuming the 
SED $f^K_\nu$ introduced above, the background radiation intensity produced 
by evolved non-merging galaxies can be calculated. When it is added to the 
background radiation spectrum produced in merging galaxies, a complete 
prediction of the background intensity between the near-infrared and 
near-ultraviolet wavebands is obtained. 
 
The background radiation intensity predicted using the 35-K model, which can 
explain the $K$- and $B$-band galaxy counts, is shown across the millimetre 
to the ultraviolet wavebands in Fig.\,12. The background intensity is in 
agreement with almost all the observed limits and detections. The only spectral 
region in which the background radiation intensity is still not very well defined 
is between about 3 and 7\,$\mu$m. At these wavelengths, the dominant source 
of background radiation switches from dust emission to starlight. There is 
almost certainly an additional, and perhaps dominant, contribution to the 
background radiation intensity in the mid-infrared waveband from very hot 
dust grains in the central regions of AGN, which are not modelled here. 
The model curves shown Fig.\,12 are not extrapolated into this region 
from the well-determined populations of galaxies at 60-$\mu$m and in the 
$K$ band.

\section{Overview of model parameters}

We have introduced a series of parameters and functions to account for 
different observations section by section through the paper. Excluding the 
world model parameters $H_0$, $\Omega_0$ and $\Omega_\Lambda$, five 
parameters are required to define the merger rate of dark matter 
haloes. Five further independent parameters [$x_0$, $p$, $z_{\rm
max}$, $(F\sigma)_0^{-1}$ and $p_\sigma$] and a SED for dusty galaxies, 
are required to fit the counts at wavelengths of 
15, 60, 175, 450 and 850\,$\mu$m and the submillimetre-wave/far-infrared 
background radiation spectrum. The most important element of the models 
are the two functions that describe the merger efficiency parameter 
$x(z)$ and the activity parameter $(F\sigma)^{-1}$. We present appropriate 
forms for these functions in equations (16) and (18) respectively, but stress 
that these forms are not unique. As more data becomes 
available, other functional forms or free-form fitting functions 
may be more appropriate. By modifying the activity parameter 
$(F\sigma)_0^{-1}$, introducing the total fraction of luminosity absorbed
by dust $A$, and including a template SED for star-forming galaxies in the 
rest frame ultraviolet waveband, the $B$-band counts can also be 
reproduced. In Table\,2 we summarize these parameters and the most 
important pieces of data used to constrain them. The values 
of the parameters required to fit the data in the 35-K model are also listed. 

\begin{table*}
\caption{A summary of the parameters and functions that have been 
introduced in this paper, generally listed in order of their first appearance. 
The values of these parameters in the best-fitting 35-K model, which 
reproduces the current redshift distribution of submillimetre-selected galaxies
adequately, are also listed. An Einstein--de Sitter world model with 
$h=0.5$ is assumed throughout. 
}
{\vskip 0.75mm}
\hrule{\vskip 1.2mm}
\begin{tabular}{ p{4.2cm} p{2.5cm} p{5.4cm} p{4.5cm} }
Name & Symbol & Value in the 35-K model & Constrained by \\
\end{tabular}
{\vskip 1.2mm}
\hrule
{\vskip 2.7mm}
\begin{tabular}{ p{4.2cm} p{2.5cm} p{5.4cm} p{4.5cm} }
Smoothed density & $\bar \rho$ & N/A & Assumed equivalent to $\Omega_0=1$\\ 
Fluctuation index & $n$ & $n \simeq -1$ & 60-$\mu$m luminosity function\\ 
& $\gamma = 1 + (n/3)$ & & \\ 
Fluctuation mass ($z=0$) & $M^*$ & $3.6 \times 10^{12}$\,M$_\odot$ &
Tully--Fisher relation \\
Merger rate constants & $\phi, \alpha$ & 1.7, 1.4 & Self-similar merging
process \\
Far-infrared SED & $f_\nu$ & N/A & 4 temperatures assumed \\
Fraction of luminous & $f_{\rm A}$ & $f_{\rm A}=0$ (generally assumed) 
& see Section 2.3\\ 
mergers powered by AGN & & & \\
Timescale of mergers & $\sigma$ & N/A & see activity parameter \\
Fraction of mergers that & $F$ & N/A & see activity parameter \\ 
yield luminous events & & & \\
\noalign{\vskip 2mm} 
Merger star-formation & $x(z)$ & 
$x_0 \simeq 10^{-4} (\gamma^2F\sigma)^{1/3}(1-f_{\rm A})$\,Gyr$^{-1/3}$ 
& Bright 60-$\mu$m counts \\
efficiency & (equation 16) & $p = 4.4$ & Far-infrared background\\
 & & $z_{\rm max} = 0.44 $ & Far-infrared background\\
\noalign{\vskip 2mm} 
Activity parameter& $(F\sigma)^{-1}$(z) & $(F\sigma)_0^{-1} = 18.8$\,Gyr$^{-1}$ 
& Bright 60-$\mu$m counts \\
(submillimeter/far-infrared)& (equation 18) & $p_\sigma \simeq 1.5$ & 
175- and 850-$\mu$m counts \\
\noalign{\vskip 2mm}
Activity parameter & & $(F\sigma)_0^{-1} = 2.5$\,Gyr$^{-1}$ & 
Faint $B$-band counts\\
($B$-band) & & $p_\sigma \simeq 1.5$ & From submillimetre results\\
\noalign{\vskip 2mm} 
Fraction of energy that& $A$ & 0.8 & Faint $B$-band counts \\ 
is reprocessed by dust& & & \\
\noalign{\vskip 1mm}
$B$- and $K$-band SEDs & $f_\nu^B$, $f_\nu^K$ & N/A & Spectral synthesis 
models \\
\noalign{\vskip 1mm}
Mass-to-light ratio of & $R_{\rm ML}$ & See equation (22) & $K$-band counts \\
evolved galaxies & & & \\
\end{tabular}
{\vskip 1.2mm}
\hrule
\end{table*}

\section{Conclusions} 

\begin{enumerate}
\item We have presented a simple model of hierarchical galaxy formation 
which incorporates the effects of obscuration by dust, in which the galaxies 
that are detected in submillimetre-wave surveys are observed during a 
merger-induced episode of star formation or AGN fuelling. The aim of this 
model is to elucidate the most important physical processes that could be at 
work in luminous dusty galaxies, rather than to provide a detailed 
quantitative description.
\item The model is constrained primarily by the intensity of background 
radiation in the far-infrared/submillimetre waveband. From these data alone, 
the luminosity density from high-redshift galaxies is inferred to exceed that 
deduced from observations in the rest frame ultraviolet and optical wavebands 
by up to an order of magnitude. The source counts and background radiation 
intensity in the submillimetre, far-, mid- and near-infrared, and optical 
wavebands are reproduced adequately in the model without introducing a 
large number of parameters. 
\item The counts of galaxies detected in the far-infrared/submillimetre and 
optical wavebands, and the associated background radiation intensities in 
these wavebands are consistent if about 4 times more energy is emitted by 
galaxies after being reprocessed into the far-infrared waveband by 
interstellar dust as is radiated directly in the optical/ultraviolet waveband. 
\item In order to account for the observed abundance of distant galaxies 
detected at 175 and 850\,$\mu$m using {\it ISO} and SCUBA, the mass-to-light 
ratio of a typical galaxy merger must decrease with redshift, by a factor of about 
10 and 200 at $z=1$ and 3 respectively. Thus high-redshift mergers must  
be typically more violent as compared with their low-redshift counterparts. We 
suggest two possible physical explanations. First, that gas is converted 
into stars/feeds an AGN uniformly more efficiently and rapidly in all merging 
galaxies as redshift increases, perhaps due to a lower bulge-to-disk ratio, 
which makes disk instabilities grow more quickly (Mihos \& Hernquist 1996). 
Secondly, that a decreasing fraction of dark matter halo mergers are associated 
with an efficient mode of star formation/AGN fuelling as redshift 
increases. 
\item In the context of galaxy formation within merging dark matter haloes, 
we have described how the physical processes that convert merging mass 
into visible radiation must evolve with redshift in order to account for the 
data in the far-infrared and submillimetre wavebands. This has previously 
been discussed by Guiderdoni et al.\ (1998), in the conventional 
context of semi-analytic models, where gas is assumed to cool into dark 
matter haloes and form stars on galactic scales. In order to account for the 
observations, an additional population of ultraluminous galaxies was 
incorporated arbitrarily into their models. We have improved our previous 
models (Blain et al.\ 1999c) significantly, by including some astrophysics 
and not simply invoking an empirical 
form of the evolution of a low-redshift luminosity function to fit 
the data. By assuming only a single population of luminous merging galaxies 
we are able to account for all the data in the far-infrared and submillimetre 
wavebands. Clear forms of evolution of both the 
efficiency with which luminosity is generated by a galaxy merger as a 
function of redshift, and of a function that connects the duration of the 
luminous phase and the fraction of dark matter halo 
mergers that generate a luminous 
event are required to reproduce the results of observations. 
The way in which gas is processed in the sub-kpc core regions of galaxy 
mergers to reproduce the necessary high efficiency and short time-scale of 
luminous events must  be investigated in future work. 
\item We find that the observed counts of both submillimetre-selected galaxies 
and Lyman-break galaxies can be accounted for in terms of merger events 
in an hierarchical model of galaxy formation, which include identical 
forms of evolution with redshift, but with different absolute normalisations. 
We find that 80\,per cent of 
the total amount of energy generated in merger-induced starbursts/AGN is 
liberated in the far-infrared waveband. It is plausible that the
submillimetre-selected galaxies and the 
Lyman-break galaxies are associated with temporally distinct phases of 
a common population of merging dark matter haloes. A scenario in which a 
short-lived, highly obscured far-infrared starburst/AGN phase 
dominates the integrated luminosity of the merger and is surrounded in time 
by a less luminous, 
more lightly obscured phase that lasts about 10 times longer 
is consistent with the data. In this scenario, the merger would be classified 
as a SCUBA galaxy if it was observed during the short-lived phase, 
and as a Lyman-break galaxy during the long-lived phase.
\item The results presented here provide excellent opportunities for 
further study. Two key scientific questions remain unanswered. First, what 
are the physical processes that are responsible for the evolution of both the 
star-formation/AGN-fuelling efficiency and the activity parameter in galaxy 
mergers as a function of redshift? Secondly, what is the relationship between 
samples of faint galaxies selected in the optical waveband and 
submillimetre-selected galaxies? Larger samples of submillimetre-selected 
galaxies and more comprehensive 
multiwaveband follow-up observations will allow these questions to be answered. 

\end{enumerate} 

\section*{Acknowledgements} 

We thank Nigel Metcalfe for providing a comprehensive list of optical count data, 
and Chris Mihos, 
Priya Natarajan, Kate Quirk, Chuck Steidel and Neil Trentham for providing 
useful comments on the manuscript. Thanks are also 
due to an anonymous referee for helpful suggestions and 
prompt reading of the manuscript. 
AWB, AJ and RJI acknowledge PPARC, IS thanks the Royal Society, and JPK 
thanks the CNRS for support. In addition, AWB thanks MENRT for support
while in Toulouse, and the Caltech 
AY visitors program for support while this work was completed.

\end{document}